\documentclass[
 reprint,
 amsmath,amssymb,
 aps,
 prb,
 superscriptaddress
]{revtex4-2}

\usepackage{float}
\usepackage[T1]{fontenc}
\usepackage{graphicx}
\usepackage{dcolumn}
\usepackage{bm}
\usepackage[colorlinks=true,allcolors=blue]{hyperref}
\usepackage{physics}
\usepackage{siunitx}
\usepackage[version=4]{mhchem}
\usepackage[usenames,dvipsnames]{xcolor}
\usepackage{enumitem}

\begin{document}

\preprint{APS/123-QED}

\title{The temperature-dependence of carrier mobility is not a reliable indicator of the dominant scattering mechanism}

\author{Alex M. Ganose}
\email{a.ganose@imperial.ac.uk}
\affiliation{Energy Technologies Area, Lawrence Berkeley National Laboratory, Berkeley, California 94720, USA}
\affiliation{Department of Chemistry, Molecular Sciences Research Hub, White City Campus, Imperial College London, Wood Lane, London, UK}

\author{Junsoo Park}
\affiliation{Energy Technologies Area, Lawrence Berkeley National Laboratory, Berkeley, California 94720, USA}

\author{Anubhav Jain}
\email{ajain@lbl.gov}
\affiliation{Energy Technologies Area, Lawrence Berkeley National Laboratory, Berkeley, California 94720, USA}

\date{\today}

\begin{abstract}
The temperature dependence of experimental charge carrier mobility is commonly used as a predictor of the dominant carrier scattering mechanism in semiconductors, particularly in thermoelectric applications.
In this work, we critically evaluate whether this practice is well founded.
A review of 47 state-of-the-art mobility calculations reveals no correlation between the major scattering mechanism and the temperature trend of mobility.
Instead, we demonstrate that the phonon frequencies are the prevailing driving forces behind the temperature dependence and can cause it to vary between $T^{-1}$ to $T^{-3}$ even for an idealised material.
To demonstrate this, we calculate the mobility of 23,000 materials and review their temperature dependence, including separating the contributions from deformation, polar, and impurity scattering mechanisms.
We conclusively demonstrate that a temperature dependence of $T^{-1.5}$ is not a reliable indicator of deformation potential scattering.
Our work highlights the potential pitfalls of predicting the major scattering type based on the experimental mobility temperature trend alone.
\end{abstract}

\maketitle

Ever since the first theories of semiconductors were developed, the temperature dependence of mobility has been used to understand the quantum behaviour of materials.
In 1931, Wilson derived expressions for charge transport in semiconductors under the assumption that lattice vibrations were the major cause of the electronic resistivity \cite{wilson1931TheoryElectronic,wilson1931TheoryElectronica}.
His work proved highly successful at predicting the temperature dependence of mobility in $n$-type germanium and lay the foundation for the modern theory of band conduction in semiconductors \cite{sommerfeld1933ElektronentheorieMetalle,debye1954ElectricalProperties}.
The temperature dependence of experimentally measured Hall mobility is now commonly used as a predictor of the dominant scattering mechanism in thermoelectric materials \cite{pei2012BandEngineering,lee2014ContrastingRole,plirdpring2012ChalcopyriteCuGaTe2}.
Knowledge of the dominant scattering mechanism is often employed to fit models of charge transport including deformation potentials and effective masses, and to obtain estimates of the optimal doping concentration and temperatures that maximise thermoelectric performance \cite{zhang2016DesigningHighperformance,ren2020EstablishingCarrier,mante2017ElectronAcoustic,liu2018HighThermoelectric,toberer2010ElectronicStructure}.

Wilson's 1931 paper demonstrated that the mobility of a system dominated by acoustic lattice scattering should exhibit a $\mu\propto T^{-1.5}$ dependence \cite{wilson1931TheoryElectronica,sommerfeld1933ElektronentheorieMetalle,bardeen1950DeformationPotentials}.
The same temperature dependence was later demonstrated for optical lattice scattering at high temperatures \cite{conwell1967high,rode1975ChapterLowField}.
Since then, a temperature dependence of $\mu\propto T^{-1.5}$ has widely been considered an experimental signature of deformation potential scattering.
Temperature trends ranging from $\mu\propto T^{-0.50}$--$T^{-0.75}$ are thought to indicate scattering due to polar optical phonons \cite{ehrenreich1957ElectronScattering,ehrenreich1959ScreeningEffects,pohls2021ExperimentalValidation}, and even more positive coefficients are ascribed to piezoelectric ($T^{-0.5}$), alloy ($T^{-0.5}$), and impurity ($T^{1.5}$) scattering \cite{krishnamurthy1985GeneralizedBrooks,rode1971ElectronTransport,yu2010FundamentalsSemiconductors}.
In all of these cases, the expected temperature trends are derived from highly-simplified models of electronic scattering in systems containing a single isotropic and parabolic band and a single dispersion-less phonon frequency.

\begin{figure}
    \centering
    \includegraphics[width=\linewidth]{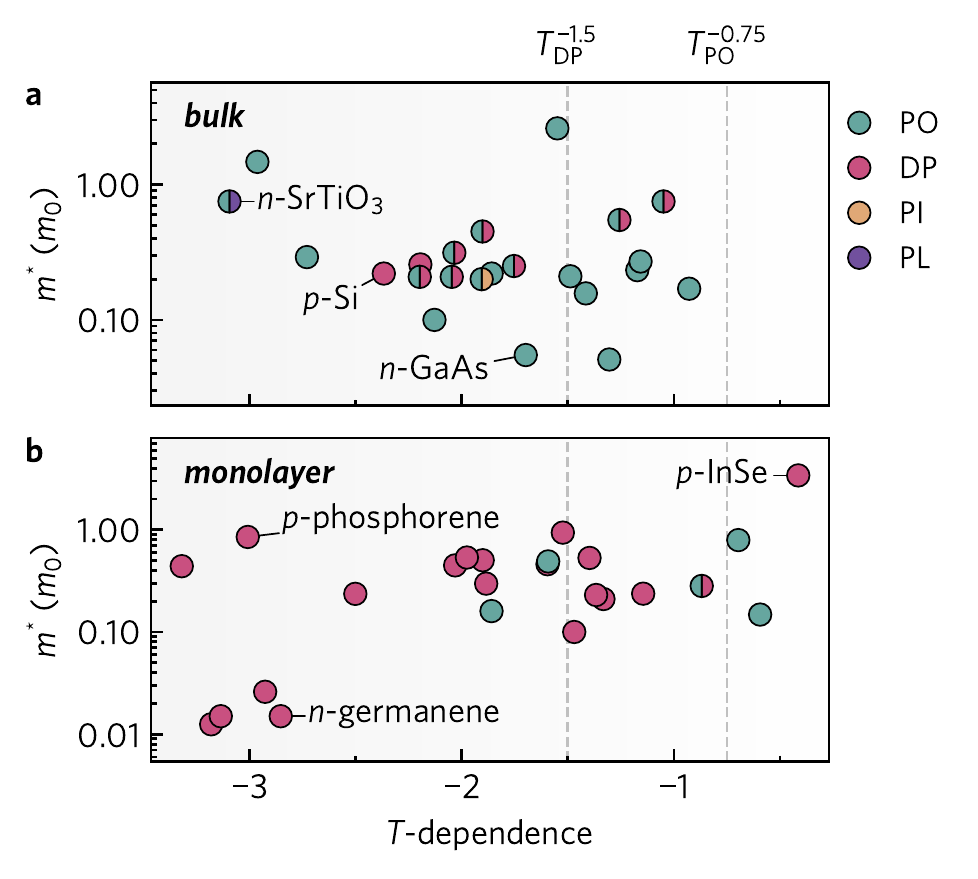}
    \caption{The temperature ($T$) dependence of mobility does not correlate with the dominant scattering mechanism. The temperature-dependence of mobility calculated using density functional perturbation theory with Wannier interpolation (DFPT+Wannier) is plotted with the dominant scattering type identified by the color of the points. There is clearly no correlation between the temperature dependence and the dominant scattering type. The temperature dependence is plotted against the band edge effective mass and separated into \textbf{a} bulk and \textbf{b} two-dimensional semiconductors to help visually clarify the points. Theoretical results were extracted from Refs.~\citep{bushick2020ElectronHole,cao2018DominantElectronphonon,ding2021ThermoelectricTransport,ge2020PhononlimitedElectronic,huang2020HighThermoelectric,lee2020InitioElectrontwophonon,li2019DimensionalCrossover,ma2018IntrinsicPhononlimited,ma2020ElectronMobility,ma2020StrainInducedUltrahigh,meng2019PhononlimitedCarrier,park2019HighThermoelectric,park2020HighThermoelectric,ponce2018PredictiveManybody,ponce2019OriginLow,su2020PhononlimitedMobility,wu2020AccuratePredictions,zhao2018IntrinsicElectronic,zhao2020AnomalousElectronic,zhou2019PredictingCharge}. The scattering types considered are polar optical (PO, teal), deformation potential (DP, pink), piezoelectric acoustic (PI, orange) and polaronic scattering (PL, purple). A particular mechanism is considered dominant if it reduces the mobility by an order of magnitude or more. We have indicated the case where two scattering mechanisms are competing as half filled circles. Only temperatures less than \SI{550}{\kelvin} were considered. See Section S1 of the Supplementary Material for the full temperature dependent results and data extraction procedure.}
    \label{fig:dfpt-wannier}
\end{figure}

State-of-the-art approaches based on density functional perturbation theory combined with Wannier interpolation (DFPT+Wannier) can now calculate the transport properties of semiconductors with predictive accuracy \cite{ponce2018PredictiveManybody,ponce2020FirstprinciplesCalculationsa,lee2020InitioElectrontwophonon}.
As the number of materials studied using DFPT+Wannier has grown --- at the time of writing this includes over 100 bulk and two-dimensional compounds --- an unexpected trend has emerged.
Many materials that were thought to be limited by acoustic deformation potential scattering based on the temperature dependence of mobility of $\mu\propto T^{-1.5}$  have instead been shown to have strong contributions from polar optical and other scattering mechanisms (Fig.~\ref{fig:dfpt-wannier}) \cite{ma2018IntrinsicPhononlimited,cao2018DominantElectronphonon}.
Accordingly, these latest computational results are challenging the commonly held assumption that the temperature dependence of mobility is a reliable indicator of the underlying scattering processes.

In this work, we critically evaluate whether the temperature dependence of carrier mobility is a reliable indicator the dominant scattering type.
We review 47 DFPT+Wannier calculations that reveals no correlation between the major scattering mechanism and the temperature trend of mobility.
Instead, we demonstrate that the phonon frequencies are largely responsible for the temperature dependence of mobility and can cause it to vary between $T^{-1}$ to $T^{-3}$ even in a simple parabolic band structure.
Finally, we extract the temperature dependence of mobility for acoustic deformation potential, polar optical, and impurity scattering in over 23,000 materials that we have calculated using the recently developed \textsc{amset} software \cite{ganose2021EfficientCalculation}.
Our results demonstrate that the temperature dependence of mobility should not be used to determine the dominant scattering mechanism.
We conclude with the potential pitfalls of assuming the dominant scattering mechanism based on the temperature dependence of mobility alone.

The temperature dependence of mobility calculated by DFPT+Wannier (23 bulk and 24 monolayer materials) shows a wide range of values spanning $-3.1$ ($n$-type \ce{SrTiO3}) to $-0.42$ (monolayer $p$-InSe) as presented in Fig.~\ref{fig:dfpt-wannier}.
We have distinguished each compound by scattering type, where a particular mechanism is considered dominant if it reduces the mobility by an order of magnitude or more.
We find no observable correlation between the dominant scattering mechanism and the temperature trend.
For example, most materials exhibit a temperature dependence more negative than $\mu\propto T^{-1.5}$.
Although this trend is commonly associated with deformation potential scattering, many of these materials are in fact dominated by polar optical phonon scattering.
Furthermore, many materials limited by deformation scattering exhibit temperature trends more negative than $\mu\propto T^{-2}$.
These results indicate that the mobility trends derived for idealised scattering in simple parabolic bands are not reliable in most materials.

We note that the majority of bulk materials are dominated by polar optical phonon scattering, whereas the monolayer materials are dominated by deformation potential scattering.
However, as most monolayers calculated using DFPT+Wannier to date have been elemental compounds that are inherently non-polar, this trend may not reveal a fundamental difference in the scattering physics between bulk and monolayer systems.

\begin{figure}
    \centering
    \includegraphics[width=0.9\linewidth]{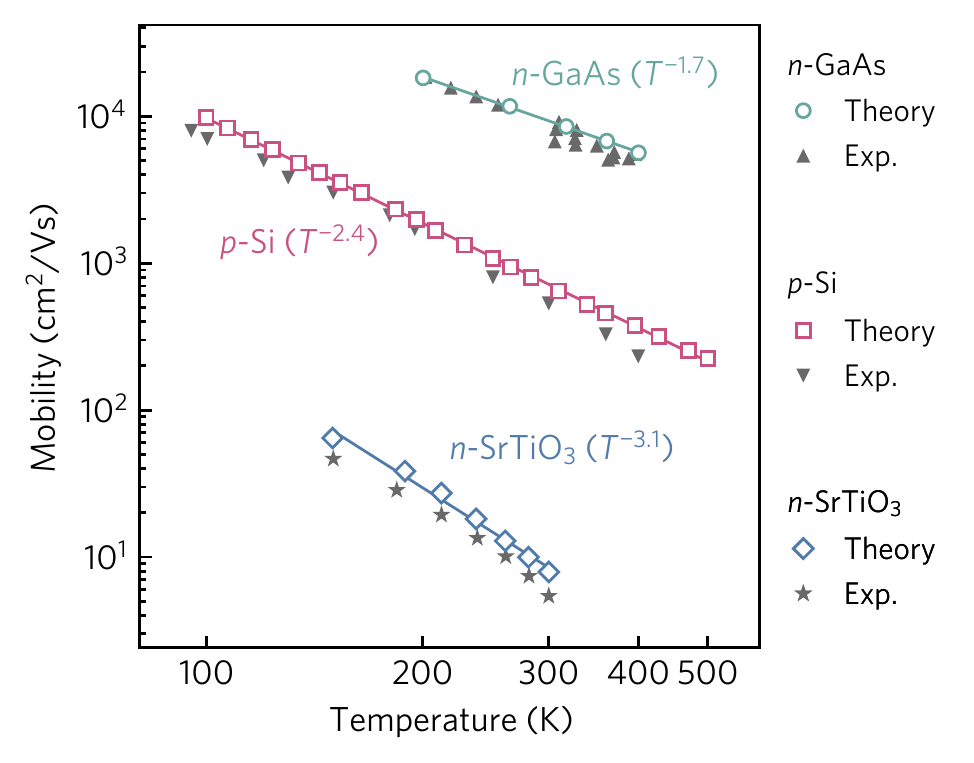}
    \caption{Electron mobility as a function of temperature for GaAs, Si, and \ce{SrTiO3}. The theoretical mobility of GaAs was calculated using density function perturbation theory combined with Wannier interpolation (DFPT+Wannier) and includes two-phonon scattering (pink squares, \cite{lee2020InitioElectrontwophonon}). The theoretical mobility of Si was obtained using DFPT+Wannier in Ref.~\cite{ponce2018PredictiveManybody}. The theoretical mobility of \ce{SrTiO3} was obtained in Ref.~\cite{zhou2019PredictingCharge} using DFPT+Wannier and a cumulant diagram-resummation approach that includes effects beyond the quasi-particle regime. The experimental mobility of GaAs (grey up triangles), Si (grey down triangles), \ce{SrTiO3} (grey starts) were obtained using Hall effect measurements from references \cite{hicks1969HighPurity,rode1971ElectronTransporta,wolfe1970ElectronMobility}, \cite{logan1960ImpurityEffects,ludwig1956DriftConductivity,morin1954ElectricalProperties,jacoboni1977ReviewChargea}, and \cite{cain2013LadopedSrTiO} respectively.}
    \label{fig:exp}
\end{figure}

\begin{figure*}
    \centering
    \includegraphics[width=0.75\linewidth]{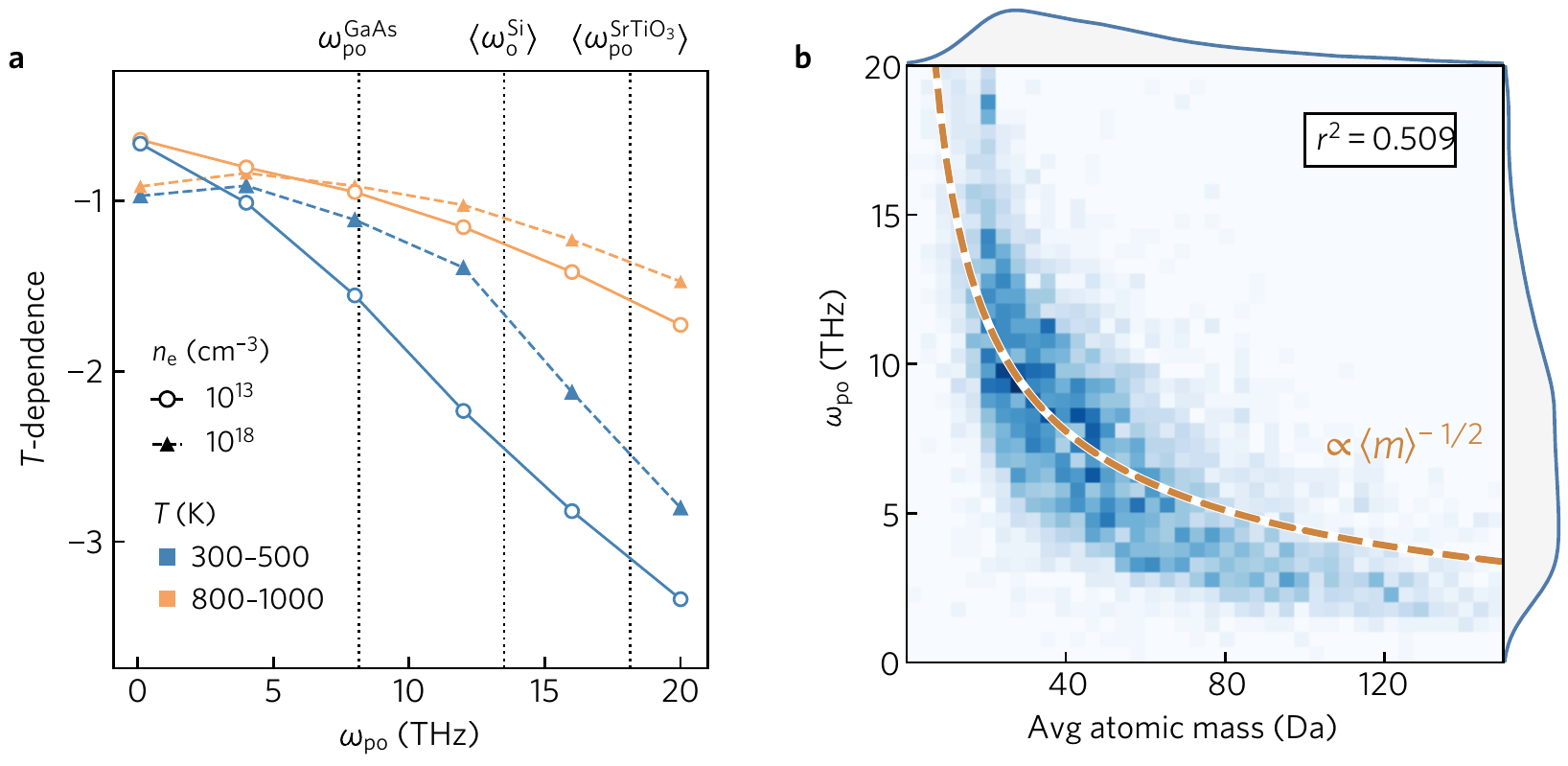}
    \caption{Polar optical phonon frequency determines the temperature dependence of mobility for transport in a single parabolic band. \textbf{a} Polar optical phonon frequency  ($\omega_\mathrm{po}$) against the temperature ($T$) dependence of mobility, calculated at low (\SI{e13}{\per\centi\meter\cubed}, open circles) and high (\SI{e18}{\per\centi\meter\cubed}, filled triangles) doping and between \SIrange{300}{500}{\kelvin} (blue) and \SIrange{800}{1000}{\kelvin} (orange). Angular brackets ($\langle \omega \rangle$) indicate averaged values. As Silicon is non-polar, we have used the average optical phonon frequency at the Brillouin zone center ($\langle \omega_\mathrm{o} \rangle$).  Calculations were performed using AMSET on a single parabolic band (effective mass of \SI{0.2}{\electronmass}) with the scattering parameters detailed in Section S3 of the Supplementary Material. \textbf{b} The atomic mass can be used as a proxy to estimate the polar optical phonon frequency. Heatmap indicating the relationship between average atomic mass and polar optical phonon frequency for all materials in the phonon frequency dataset (generated from density functional theory calculations; machine learning predictions are excluded). The phonon frequency is roughly proportional to the inverse square root of the averaged mass ($\omega_\mathrm{po} \propto \left < m \right >^{-1/2}$), as can be derived directly from the phonon dynamical matrix. Darker points indicate more materials. The $r^2$ correlation coefficient in the white box indicates reasonable correlation.}
    \label{fig:pop}
\end{figure*}

\begin{figure*}
    \centering
    \includegraphics[width=\linewidth]{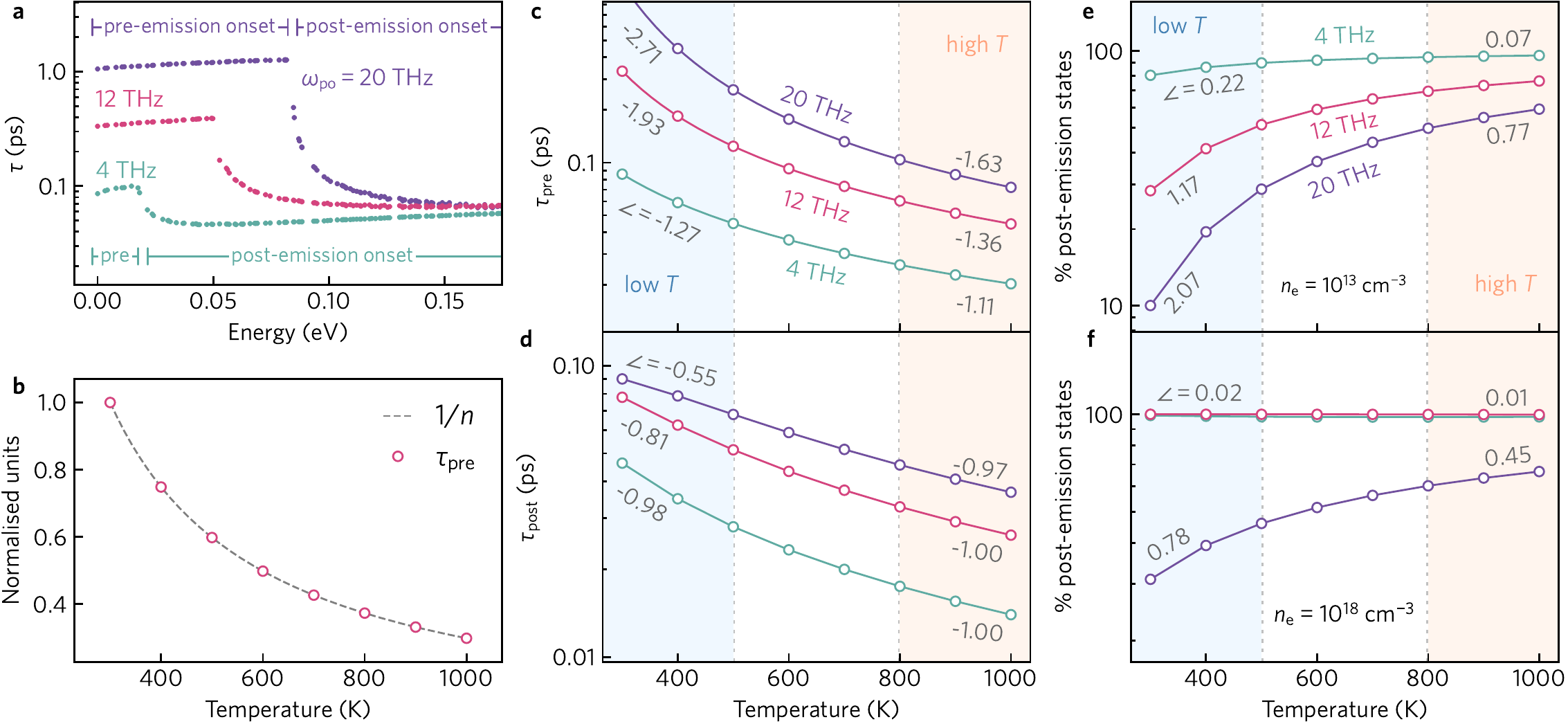}
    \caption{The temperature dependence of the electron lifetimes are determined by the Bose--Einstein phonon occupation number ($n$) which is the main factor that controls the temperature dependence of mobility. \textbf{a} Energy dependence of the electron lifetimes ($\tau$) for different polar optical phonon frequencies, indicating pre-phonon emission and post-emission regimes. Larger frequencies increase the pre-emission onset energy range. \textbf{b} Temperature against the pre-phonon emission electron lifetimes ($\tau_\mathrm{pre}$) and the inverse phonon occupation ($1/n$). At higher temperatures, the phonon occupation increases and the lifetime is shortened. The temperature dependence of $n$ directly controls the temperature dependence of $\tau_\mathrm{pre}$. The phonon occupation and scattering rate are normalised to their values at \SI{300}{\kelvin}. \textbf{c} Pre- and \textbf{d} post-emission lifetimes, calculated at a carrier concentration of \SI{e13}{\per\centi\meter\cubed}. The temperature dependence of the lifetimes is stronger (more negative) for pre-emission states and at lower temperatures. Percentage of states that contribute to mobility in the post-emission energy range against temperature for \textbf{e} low doping (\SI{e13}{\per\centi\meter\cubed}) and \textbf{f} high doping (\SI{e18}{\per\centi\meter\cubed}) concentrations (see text for more details). The percentage of post-emission states is lower at low temperatures and low doping. In \textbf{c}-\textbf{f}, grey numbers indicate the temperature dependence of the associated curve. Calculations were performed using AMSET with scattering parameters detailed in Section S3 of the Supplementary Material.}
    \label{fig:dep_mech}
\end{figure*}

We stress that DFPT+Wannier is a state-of-the-art approach that can yield excellent agreement with experimental Hall effect measurements.
This is highlighted in Fig.~\ref{fig:exp} where we present the calculated and experimental carrier mobilities of $n$-GaAs, $p$-Si, and $n$-\ce{SrTiO3}.
For each material, the theoretical mobility exhibits excellent agreement with both the magnitude and temperature dependence of the experimental measurements.
However, in each case the observed temperature trend differs from the trend predicted by the dominant scattering mechanism.
For example, $p$-Si exhibits a $\mu \propto T^{-2.4}$, despite being dominated by deformation potential scattering, a very large deviation from the nominal $T^{-1.5}$ dependence.
In the following section we examine these materials in more detail with the goal of uncovering why a given scattering type can exhibit a wide range of temperature trends. 

GaAs is a classic zinc-blende semiconductor with an isotropic conduction band pocket centered at the $\Gamma$-point and scattering dominated by polar optical phonons \cite{lee2020InitioElectrontwophonon}.
Despite its simple band structure and essentially single source of scattering, GaAs exhibits a $\mu\propto T^{-1.7}$ dependence that is very close to that associated with deformation potential scattering.
As we shall demonstrate, the value of $-1.7$ is not intrinsic to the dominant scattering mechanism itself but is instead a consequence of the physical properties of GaAs, in particular: (i) its large polar optical phonon frequency and (ii) slight non-parabolicity in the conduction band.

To investigate further, we performed mobility calculations for a single parabolic band with an effective mass $m^{*}_e$ = \SI{0.2}{\electronmass} using the \textsc{amset} package \cite{ganose2021EfficientCalculation} and only included polar optical scattering (known to dominate in GaAs).
\textsc{amset} has been shown provide scattering rates and mobility within $\sim$ \SI{10}{\percent} of DFPT+Wannier when benchmarked on 23 semiconductors \cite{ganose2021EfficientCalculation}.
Further details on the \textsc{amset} methodology and the calculation procedure are given in Section S3 of the Supplementary Material.

Our transport calculations reveal that systems with smaller phonon frequencies will show a weaker (less negative) temperature dependence of mobility for a fixed temperature range.
Indeed, simply adjusting the polar optical phonon frequency can cause the mobility to decay as gradually as $\mu\propto T^{-0.67}$ ($\omega_{\text{po}}$ = \SI{0.1}{\tera\hertz}) or as rapidly as $\mu\propto T^{-3.34}$ (\SI{20}{\tera\hertz}) for a single parabolic band at low temperature and doping ($T$ = \SIrange{300}{500}{\kelvin}; $n$ = \SI{e13}{\per\cubic\cm}, Fig.~\ref{fig:pop}a, solid blue line).
We note that the temperature dependence for small $\omega_{\text{po}}$ falls within the range of values broadly associated polar optical phonon scattering ($T^{-0.50}$--$T^{-0.75}$).
When calculations are performed using the polar optical frequency of GaAs ($\omega_{\text{po}}$ = \SI{8.16}{\tera\hertz}), the mobility exhibits a dependence of $\mu\propto T^{-1.58}$ close to the experimental trend of $\mu\propto T^{-1.7}$.
Our analysis indicates that simple compounds composed of light atoms with high-frequency optical modes are likely to exhibit a more negative $T$-dependence of mobility than compounds composed of heavy atoms which generally exhibit low phonon frequencies (as demonstrated in Fig.~\ref{fig:pop}b which reveals the inverse relationship between atomic mass and $\omega_\mathrm{po}$).

We note that the temperature dependence of mobility also depends on the temperature and the doping concentration.
At higher temperatures, the temperature dependence is weakened (made less negative).
For example, at a phonon frequency of \SI{20}{\tera\hertz} the mobility dependence is reduced from $T^{-3.34}$ between \SIrange{300}{500}{\kelvin} to $T^{-1.81}$ between \SIrange{800}{1000}{\kelvin} (Fig.~\ref{fig:pop}a, solid orange line). 
Thermoelectric materials generally exhibit optimal performance at degenerate or near degenerate doping (termed ``high doping'').
A higher doping concentration results in a weakening of the temperature dependence of mobility across all phonon frequencies (Fig.~\ref{fig:pop}a, dashed blue line, $n_\mathrm{e} = \SI{e18}{\per\cubic\centi\meter}$).
At high temperature and high doping, the mobility exhibits the weakest temperature dependence and does not become more negative than $\mu \propto T^{-1.50}$ even at the largest phonon frequencies (Fig.~\ref{fig:pop}a, $\omega_\mathrm{po} = \SI{20}{\tera\hertz}$, $n_\mathrm{e} = $\SI{e18}{\per\cubic\centi\meter}, $T$ = \SIrange{800}{1000}{\kelvin}).
Note, however, that even at high-temperature, high-doping conditions, mobility limited by polar-optical scattering can exhibit a similar temperature dependence as that broadly associated with lattice deformation potential scattering ($T^{-1.50}$).
In Section S2 of the Supplementary Material, we explicitly investigate the impact of changing the temperature and carrier concentration on the temperature dependence of mobility, and confirm the trends discussed above.

\begin{figure*}
    \centering
    \raisebox{-0.5\height}{\includegraphics[width=0.65\linewidth]{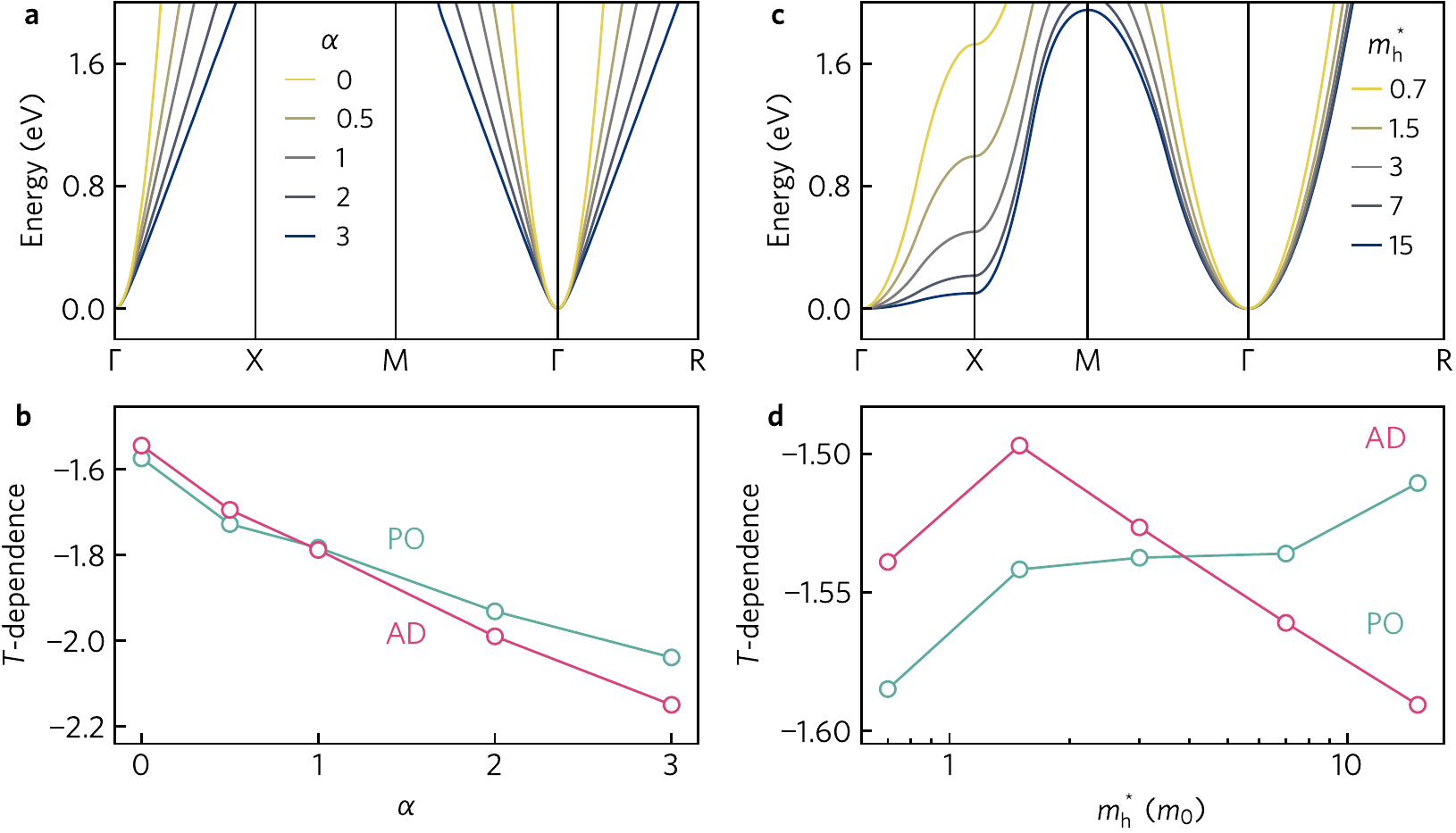}}
    \hfill
    \raisebox{-0.45\height}{\includegraphics[width=0.294\linewidth]{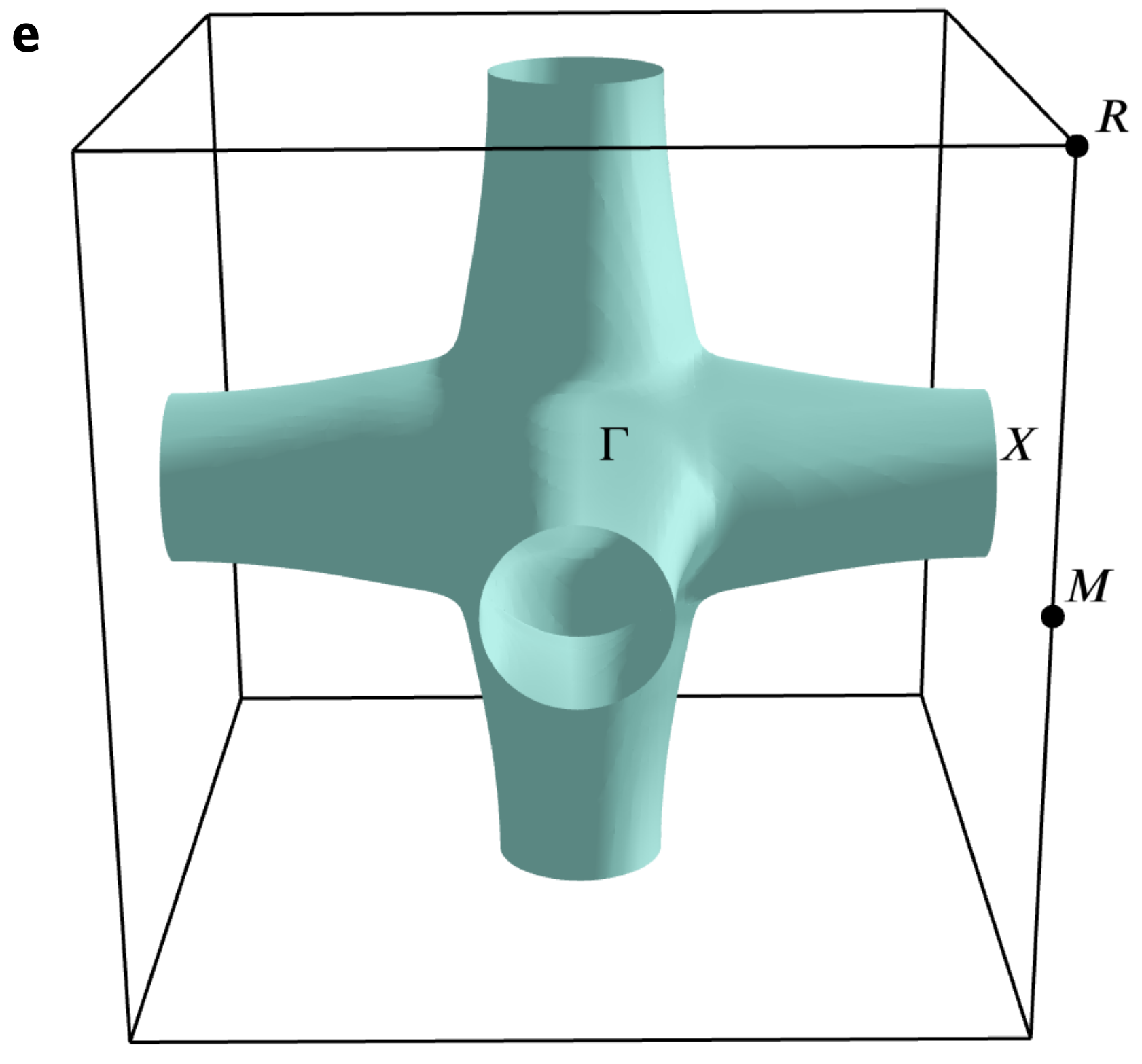}}
    \caption{The electronic band structure can control the temperature dependence of mobility. Effect of Kane non-parabolicity parameter ($\alpha$) on \textbf{a} the electronic band structure and \textbf{b} the temperature ($T$) dependence of mobility calculated using polar optical phonon scattering (PO) and acoustic deformation potential scattering (AD). Effect of a heavy anisotropic effective mass ($m_\mathrm{h}^{*}$) on the \textbf{c} electronic band structure and \textbf{d} temperature dependence of mobility. \textbf{e} Fermi surface of the anisotropic effective mass where $m_\mathrm{h}^{*}$ = \SI{7}{\electronmass} at \SI{0.4}{\electronvolt} above the valence band edge. Fermi surface visualized using the \textsc{ifermi} package \cite{ganose2021IFermiPython}.}
    \label{fig:bandstructure}
\end{figure*}

The relationship between phonon frequency and the temperature dependence of mobility arises due to the temperature dependence of the electron lifetimes ($\tau$).
The temperature dependence of the electron lifetimes in turn results from the Bose--Einstein occupation factor of the phonons $n = 1 / [ \exp (\hbar \omega / k_\mathrm{B} T) - 1]$, where $\hbar$ is the reduced Planck's constant and $k_\mathrm{B}$ is the Boltzmann constant.
Greater phonon occupation results in decreased electron lifetimes.
For example, Fig.~\ref{fig:dep_mech}b reveals the lifetimes of the low energy electronic states (i.e., pre-phonon emission scattering, at energies less than $\omega_\mathrm{po}$ from the conduction band minimum, Fig.~\ref{fig:dep_mech}a) are inversely proportional to the Bose--Einstein phonon occupation.
At low temperatures, the phonons will not be sufficiently excited and their population will increase exponentially with temperature, thereby rapidly decreasing the electron lifetimes (see blue region in Fig.~\ref{fig:dep_mech}c) and hence the electron mobility.
At higher temperatures, phonon occupation increases roughly linearly with temperature resulting in more gradual decay of electron lifetimes (see orange region in Fig.~\ref{fig:dep_mech}c).
The temperature at which the rate of occupation transitions from exponential to linear increase is determined by the phonon frequency.
A low frequency ($\hbar\omega < k_\mathrm{B}T$) means the phonon occupation increases more linearly and hence the mobility will show a weaker temperature dependence (see teal line [\SI{4}{\tera\hertz}] in Fig.~\ref{fig:dep_mech}c).
A higher frequency ($\hbar\omega > k_\mathrm{B}T$) means the phonon occupation increases more exponentially with temperature (see purple line [\SI{20}{\tera\hertz}] in Fig.~\ref{fig:dep_mech}c).
Accordingly, for a fixed temperature range, the temperature dependence of mobility is less negative in systems with smaller phonon frequencies, exactly as revealed in Fig.~\ref{fig:pop}a.

The weakening in the temperature dependence of mobility at higher temperatures is also explained by the more linear change of phonon occupation in the high temperature regime.
There is an additional effect, arising from the ratio of pre- and post-emission onset lifetimes, that reinforces the impact of phonon occupations. This is discussed further in Section S4 of the Supplementary Material.

At higher doping concentrations, the temperature dependence of mobility is also weakened (made less negative).
This is because increased doping activates higher energy electronic states (after the emission scattering onset, see pink lines [\SI{12}{\tera\hertz}] in Figs.~\ref{fig:dep_mech}e for low doping and \ref{fig:dep_mech}f  for high doping) whose lifetimes have much weaker temperature dependence.
This can be seen in the blue region of Fig.~\ref{fig:dep_mech}d, where the lifetimes of the post-emission electronic states show dramatically reduced temperature dependence (between $\tau \propto T^{-0.98}$--$T^{-0.81}$) compared to the lower energy pre-emission states (Fig.~\ref{fig:dep_mech}c, $\tau \propto T^{-1.27}$--$T^{-2.71}$).

Although thus far we have restricted our analysis to polar optical phonon scattering, the same factors will also determine the temperature dependence of systems limited by optical deformation potential scattering, albeit with some caveats.
The major complicating factor is that in polar materials, scattering occurs only by polar longitudinal-optical modes near the zone center. 
Accordingly, often only a few phonon frequencies control the entire scattering, and even just one frequency in the simplest of polar systems.
However, in optical deformation potential scattering, both longitudinal and transverse modes across the full Brillouin zone can scatter carriers, leading to a wide spectrum of phonon frequencies that contribute to scattering.
Regardless, the overall trends discussed above are expected to hold for any systems dominated by electron-phonon interactions.

\begin{figure*}
    \centering
    \includegraphics[width=0.7\linewidth]{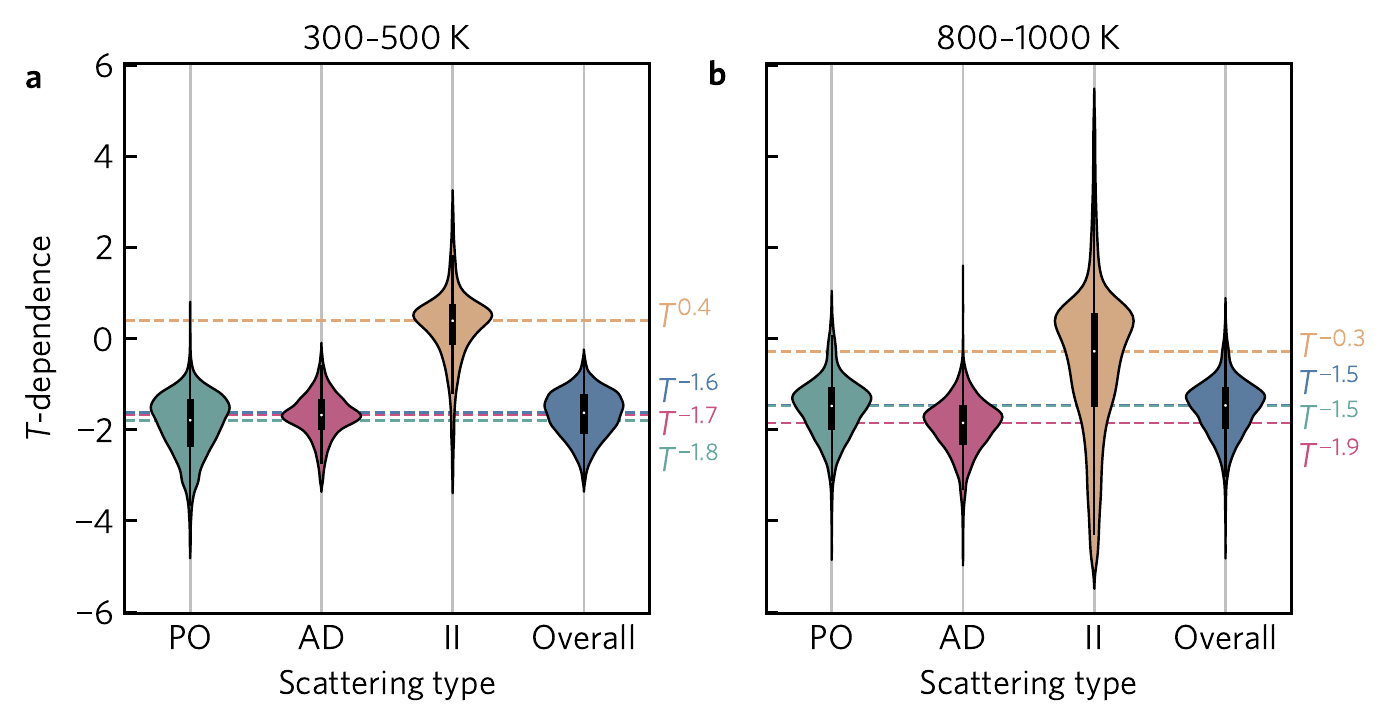}
    \caption{The temperature dependence of mobility varies widely even among a single scattering mechanism. Histograms of the temperature ($T$) dependence of mobility for different scattering mechanisms at \textbf{a} low (\SIrange{300}{500}{\kelvin}) and \textbf{b} high (\SIrange{800}{1000}{\kelvin}) temperatures obtained from 23,000 materials. Calculations were performed using the \textsc{amset} package using a carrier concentration of \SI{e17}{\per\cubic\centi\meter}) and are separated into polar optical (PO), acoustic deformation potential (AD), and ionized impurity (II) scattering. The ``Overall'' distribution gives the temperature dependence using the total scattering rate ($\tau^{-1}$) obtained from Matthiessen's rule ($\tau^{-1}_\mathrm{overall} = \tau^{-1}_\mathrm{po} + \tau^{-1}_\mathrm{ad} + \tau^{-1}_\mathrm{ii}$).}
    \label{fig:amset}
\end{figure*}

With this in mind, we return to the remaining systems previously highlighted, $n$-\ce{SrTiO3} and $p$-Si.
In each case, the average phonon frequency (polar frequency for \ce{SrTiO3}) determines the temperature dependence of mobility. 
In SrTiO$_{3}$, the mobility is limited by a combination of polar optical and soft-ferroelectric phonons and polaronic effects \cite{zhou2019PredictingCharge}.
The mobility dependence of $\mu\propto T^{-3.1}$ is one of the most negative in our dataset, and is dominated by two optical phonons, with frequencies \SI{13.3}{\tera\hertz} and \SI{23.0}{\tera\hertz} \cite{zhou2019PredictingCharge}.
Using the arithmetic mean of these phonon frequencies (\SI{18.2}{\tera\hertz}) leads to a mobility dependence around $\mu\propto T^{-3}$ as demonstrated in Fig.~\ref{fig:pop}a, very close to the experimental value.
The mobility of $p$-type Si  exhibits a temperature dependence of $\mu\propto T^{-2.4}$ and is dominated by optical deformation scattering from phonons with frequencies between \SIrange{12}{15}{\tera\hertz} \cite{li2021deformation}.
Using the average of the optical mode frequencies (\SI{13.5}{\tera\hertz}) gives rise to an expected mobility dependence of $\mu \propto T^{-2.5}$, as illustrated in Fig.~\ref{fig:pop}a.
This is in excellent agreement with the experimental trend even though Si is not limited by polar optical phonon scattering, but rather deformation potential scattering.
Accordingly, the temperature dependence of mobility in GaAs, Si, and \ce{SrTiO3} is controlled almost entirely by the phonon frequencies irrespective of the dominant scattering mechanisms.

It is important to stress that the temperature dependence of mobility also depends on features in the electronic band structure.
For example, although GaAs possesses a single isotropic band, it is not perfectly parabolic away from the band edge.
By increasing the non-linearity in a single isotropic band quantified by the Kane parameter ($\alpha$), the temperature-dependence of mobility can vary from $\mu\propto T^{-1.58}$ in the fully parabolic case ($\alpha=0$) to $\mu\propto T^{-2.04}$ (polar optical limited) and $T^{-2.15}$ (acoustic deformation limited) in the most non-parabolic band structure explored ($\alpha=3$).
Accordingly, even a simple band structure feature in a single band (without considering degeneracy or multiple band pockets) is capable of modulating the temperature trend of mobility by up to \SI{40}{\percent} (Fig.~\ref{fig:bandstructure}b).
Additionally, the $T$-dependence of polar optical phonon limited and acoustic deformation potential limited mobility is similar (within \SI{5}{\percent} between the independent mechanisms) across the full range of $\alpha$ values, further highlighting the difficulty of assigning a particular scattering mechanism based on the $T$-dependence of mobility alone.
In our calculations, all scattering parameters are fixed except that the degree of parabolicity is modulated using the Kane parameter $\alpha$ (see Section S3 of the Supplementary Material).

We also find that band anisotropy has a small but non-negligible effect on the temperature dependence (Fig.~\ref{fig:bandstructure}c and d), with the mobility dependence varying by $\sim$\SI{8}{\percent} (polar optical) and \SI{6}{\percent} (acoustic deformation) as anisotropy increases (details on the generation and calculation of anisotropic band structures are given in Section S3 of the Supplementary Material).
In Section S3 of the Supplementary Material, we demonstrate that ionized impurity scattering exhibits a similar modulation with non-parabolicity and that these effects also hold at larger doping and temperature regimes and will still play a crucial role in determining the temperature dependence at typical thermoelectric doping levels.

In summary, the following trends are expected to hold in any systems dominated by electron-phonon scattering:
\begin{itemize}[align = left]
  \item[\textbf{Observation 1}] Compounds containing lighter elements should generally exhibit stronger (more negative) temperature dependence due to high $\omega_\mathrm{o}$ (as demonstrated in Fig.~\ref{fig:pop}b).
  \item[\textbf{Observation 2}] The temperature dependence of mobility should weaken (become less negative) at higher temperatures.
  \item[\textbf{Observation 3}] The temperature dependence of mobility should weaken with doping (barring heavily degenerate doping).
  \item[\textbf{Observation 4}] Band structure features can modulate the temperature dependence from the idealised value (typically making it more negative) by up to $\sim$\SI{40}{\percent}. E.g., from $\mu \propto T^{-1.58}$ to $T^{-2.04}$ for the case of non-parabolic bands with polar optical phonon scattering.
\end{itemize}

\begin{figure*}
    \centering
    \includegraphics[width=0.7\linewidth]{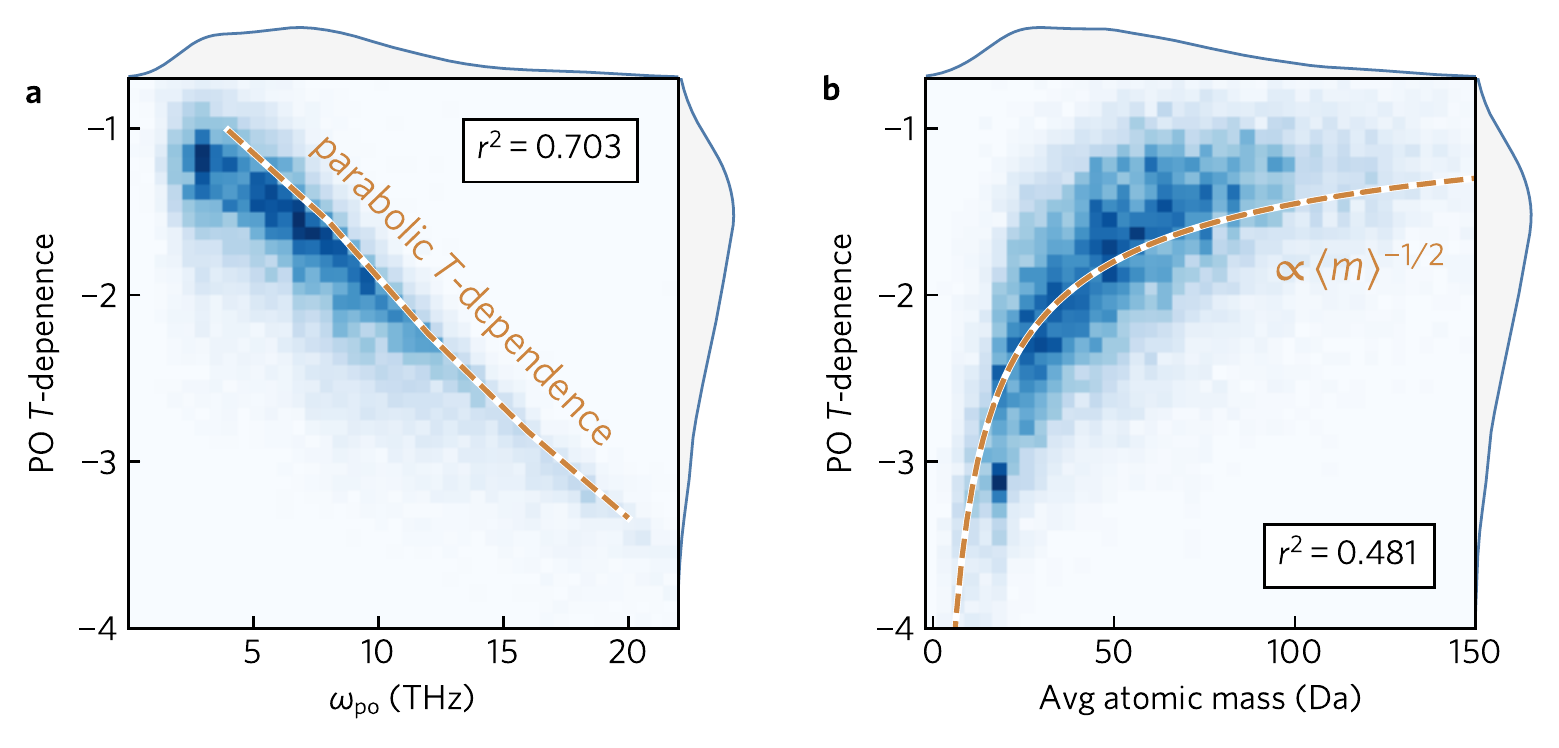}
    \caption{The atomic mass can be used as a proxy to estimate the temperature dependence of mobility limited by optical phonon scattering. \textbf{a} Heatmap indicating a linear correlation between the polar optical phonon frequency ($\omega_\mathrm{po}$) and the temperature dependence of mobility (in the range \SIrange{300}{500}{\kelvin}) limited by polar optical phonon scattering. The $T$-dependence for a parabolic band under low doping conditions taken from Fig.~\ref{fig:pop}a is superimposed as a dashed grey line. \textbf{b} Heatmap indicating the correlation between average atomic mass and the temperature dependence of mobility limited by optical phonon scattering. The $r^2$ correlation coefficients in the white boxes indicate reasonable correlation for each trend. The temperature dependence is measured in the range $T =$ \SIrange{300}{500}{\kelvin}. Darker points indicate more materials.}
    \label{fig:hist_trends}
\end{figure*}

The wide range in the temperature dependence of mobility due to phonon frequency and band structure features clearly indicates that a particular scattering mechanism must not be associated with a nominal $T$-dependence, even at high-doping, high-temperature conditions where it is usually deemed safer to do so.
This is highlighted in Fig.~\ref{fig:amset}a, where we present the temperature dependence of 23,000 materials calculated high-throughput using the \textsc{amset} package \cite{ganose2021EfficientCalculation} at both $n$- and $p$-type doping conditions (\SI{e17}{\per\cubic\centi\meter}).
The transport dataset was generated using band structures obtained from density functional theory and materials parameters (deformation potentials, elastic and dielectric constants, and phonon frequencies) from machine learning models.
As our goal is to simulate a realistic distribution of materials parameters rather than the exact material parameters for specific materials, strict accuracy in the inputs is not essential.
Regardless, in Section S5 the Supplementary Material, we validate our machine learning models (generated using \textsc{automatminer}\cite{dunn2020BenchmarkingMaterials}) using cross validation studies.
One benefit of \textsc{amset} is that the overall scattering rate can be separated into the contributions from polar optical phonons, acoustic deformation potentials, and ionized impurities, and the mobility calculated separately for each scattering mechanism.
We note that \textsc{amset} does not yet implement optical deformation potential scattering, but as previously mentioned, its effect is likely intermediate between polar-optical scattering (owing to inelasticity) and acoustic deformation potential scattering (owing to the participation of phonons across the entire Brillouin zone).
Further details on the generation of the \textsc{amset} dataset is provided in Section S5 of the Supplementary Material.

The average $T$-dependence of mobility considering all scattering mechanisms (where the overall scattering rate for a given material is obtained from Matthiessen's rule as $\tau^{-1}_\mathrm{overall} = \tau^{-1}_\mathrm{po} + \tau^{-1}_\mathrm{ad} + \tau^{-1}_\mathrm{ii}$ and the subscripts refer to polar optical phonon [PO], acoustic deformation potential [AD] and ionized impurity [II] scattering) is found to be $\mu \propto T^{-1.6}$ at low temperature (\SIrange{300}{500}{\kelvin}) and $\mu \propto T^{-1.5}$ at high temperature (\SIrange{800}{1000}{\kelvin}).
When considering only polar optical scattering at low temperatures, the average $T$-dependence is approximately $\mu \propto T^{-1.8}$ and the broad interquartile range (containing the middle \SI{50}{\percent} of the distribution) from $T^{-1.4}$ to $T^{-2.3}$ highlights the significant variation across materials for a single scattering type.
The distribution of polar optical scattering overlaps considerably with acoustic deformation potential scattering, which has an average of $\mu \propto T^{-1.5}$ but a tighter interquartile range from $T^{-1.4}$ to $T^{-1.9}$.
The average temperature dependence of ionized impurity scattering limited mobility is found to be $\mu \propto T^{0.4}$ . 
This weakly positive temperature dependence (indicating that mobility weakly increases with temperature, Fig.~\ref{fig:amset}a) can be rationalized by considering that at high temperatures the Thomas--Fermi screening length will increase and shield more electronic states from charge centers.
In the high-temperature regime, the average polar optical phonon $T$-dependence is reduced to $\mu \propto T^{-1.5}$ and the interquartile range narrows to between $T^{-1.1}$--$T^{-1.9}$.
This is in agreement with our previous analysis that predicted weaker temperature dependence at higher temperatures.
In this regime, acoustic deformation potential scattering limited mobility has an average trend of $T^{-1.9}$ which is again close to that from polar optical phonon scattering. 
In Section S6 of the Supplementary Material, we demonstrate that the same behaviour is also observed at a higher doping concentrations. The broad overlap of polar optical and acoustic deformation potential distributions, even at high temperatures, and nearly indistinguishable mean trends, indicate that identifying the dominant scattering mechanism through the temperature dependence of mobility alone will be highly unreliable.
This corroborates the trends of the independent but considerably smaller DFPT+Wannier dataset and highlights the pitfalls of using mobility trends devised for idealised scattering in parabolic band structures.
In contrast, we note that impurity scattering is distinguishable from the phonon-mediated scattering mechanisms due to its positive temperature dependence.

In Fig.~\ref{fig:hist_trends}a, we assess whether our prediction that larger polar optical phonon frequencies correspond to stronger $T$-dependence of mobility holds across the \textsc{amset} dataset.
The results are illustrated in the form of a heatmap, where darker colours indicate greater density of materials.
There is a clear inverse relationship between polar phonon frequency and the $T$-dependence of mobility, as confirmed by the $r^2$ correlation coefficient of 0.70.
Accordingly, the behaviour of the $T$-dependence of mobility with polar optical phonon frequency is observed across a wide range of band structure types beyond the simple parabolic case discussed previously.
We note that the distribution of $T$-dependencies broadly agrees with the single parabolic band model presented in Fig.~\ref{fig:pop}a.
The parabolic band results provides an upper bound for the temperature dependencies seen from our dataset (we outline a potential origin of this behaviour in Section S7 of the Supplementary Material).
For example, a polar optical frequency of \SI{10}{\tera\hertz} corresponds to a $\mu\propto T^{-1.9}$ in a parabolic band structure, whereas the same phonon frequency gives an average dependence of $T^{-2.1}$ in the full \textsc{amset} dataset.
There is also considerable width to the distribution at a single value of $\omega_\mathrm{po}$.
For example, at $\omega_\mathrm{po}$ = \SI{10}{\tera\hertz} the $T$-dependence varies from $-3.1$ to \SI{-1.5}{\tera\hertz} depending on the material.
In Section S8 of the Supplementary Material we investigate whether common band structure features (such as band edge degeneracy, effective mass, Fermi surface complexity factor \cite{gibbsEffectiveMassFermi2017}, and band width) can explain the additional variation in temperature dependence at a particular $\omega_\mathrm{po}$.
We find that no single feature or combination of features can explain the additional variation and therefore posit that the $T$-dependence must be controlled by a complex interplay between band structure effects, wave function overlaps, and the behaviour of the scattering rates across the band structure.

Noticeably, we find that the average atomic mass exhibits a reasonable correlation ($r^2=0.48$) to the $T$–dependence of mobility limited by polar optical phonon scattering (Fig.~\ref{fig:hist_trends}b).
This arises due to the relationship between average atomic mass and polar phonon frequency revealed in Fig.~\ref{fig:pop}b.
Accordingly, the composition alone can be used as as a rough guide to the temperature dependence of mobility, without needing to know the phonon frequencies (which typically requires first principles calculations or neutron diffraction measurements).

Lastly, we briefly discuss the potential consequences of inferring the dominant scattering mechanism from the temperature dependence of experimental mobility alone.
Most commonly, the major scattering mechanism is used to optimize and understand thermoelectric materials.
For example, a $T$-dependence of $\mu \propto T^{-1.5}$ is often assumed to indicate deformation potential scattering and has led many researchers to fit oversized deformation potentials in an effort to reproduce the experimental mobility.
Egregious examples include SnSe \cite{chen2014ThermoelectricProperties}, PbTe \cite{wang2013HighThermoelectrica}, and \ce{BiCuSeO} \cite{fan2017UnderstandingElectronic}, in which the mobility dependence of $T^{-1.5}$ has lead to enormous fitted deformation potentials of \SI{24}{\electronvolt} (SnSe and BiCuSeO) and \SI{22}{\electronvolt} (PbTe), over \SIrange{10}{20}{\electronvolt} larger than those calculated from density functional theory \cite{ganose2021EfficientCalculation,zhao2020PredictionHigher}.
Rather, first-principles scattering rate calculations recently revealed that polar optical phonon scattering is approximately an order of magnitude stronger than deformation potential scattering in each of these materials and dictates both the magnitude and temperature dependence of mobility \cite{ma2018IntrinsicPhononlimited,cao2018DominantElectronphonon,zhao2020PredictionHigher}.
The dominant scattering mechanism also controls the optimal doping and temperature necessary to maximise the thermoelectric figure of merit $zT$ \cite{pohls2021ExperimentalValidation}. 
Accordingly, incorrect identification of the primary scattering processes can stymie efforts to engineer greater performance.

In conclusion, we demonstrated that the temperature dependence of mobility does not correlate with the dominant scattering mechanism as previously assumed.
Instead, the phonon frequencies are the major factor that control the temperature dependence.
For systems dominated by electron-phonon coupling, we expect the following trends to hold: i) materials with lighter elements should exhibit stronger (more negative) temperature dependence due to their larger phonon frequencies; ii) the temperature dependence of mobility should weaken at higher temperatures and larger doping concentrations; and iii) band structure features such as non-parabolicity can cause a reasonable modulation of the temperature dependence.
These trends hold across simple parabolic band structures and a dataset of 23,000 materials covering a diverse range of band structure and scattering properties.
Our work conclusively demonstrates that a temperature dependence of $T^{-1.5}$ is not a reliable indicator of deformation potential scattering and assuming as such can lead to a flawed understanding of electron transport and optimal thermoelectric performance.


\section*{Acknowledgements}

The authors thank Xin Chen for helpful discussions. The authors thank Hao Zhang for graciously sharing additional supplementary data files. This work was funded and intellectually led by the U.S. Department of Energy (DOE) Basic Energy Sciences (BES) program --- the Materials Project --- under grant no. KC23MP. This research used resources of the National Energy Research Scientific Computing Center, which is supported by the Office of Science of the U.S. Department of Energy under Contract no. DEAC02-05CH11231. Lawrence Berkeley National Laboratory is funded by the DOE under award DE-AC02- 05CH11231.

\bibliography{refs}

\end{document}


\section{DFPT+Wannier dataset}

\begin{figure}[p]
  \vspace*{-4.5cm}
  \makebox[\linewidth]{
  \includegraphics[width=1.3\linewidth]{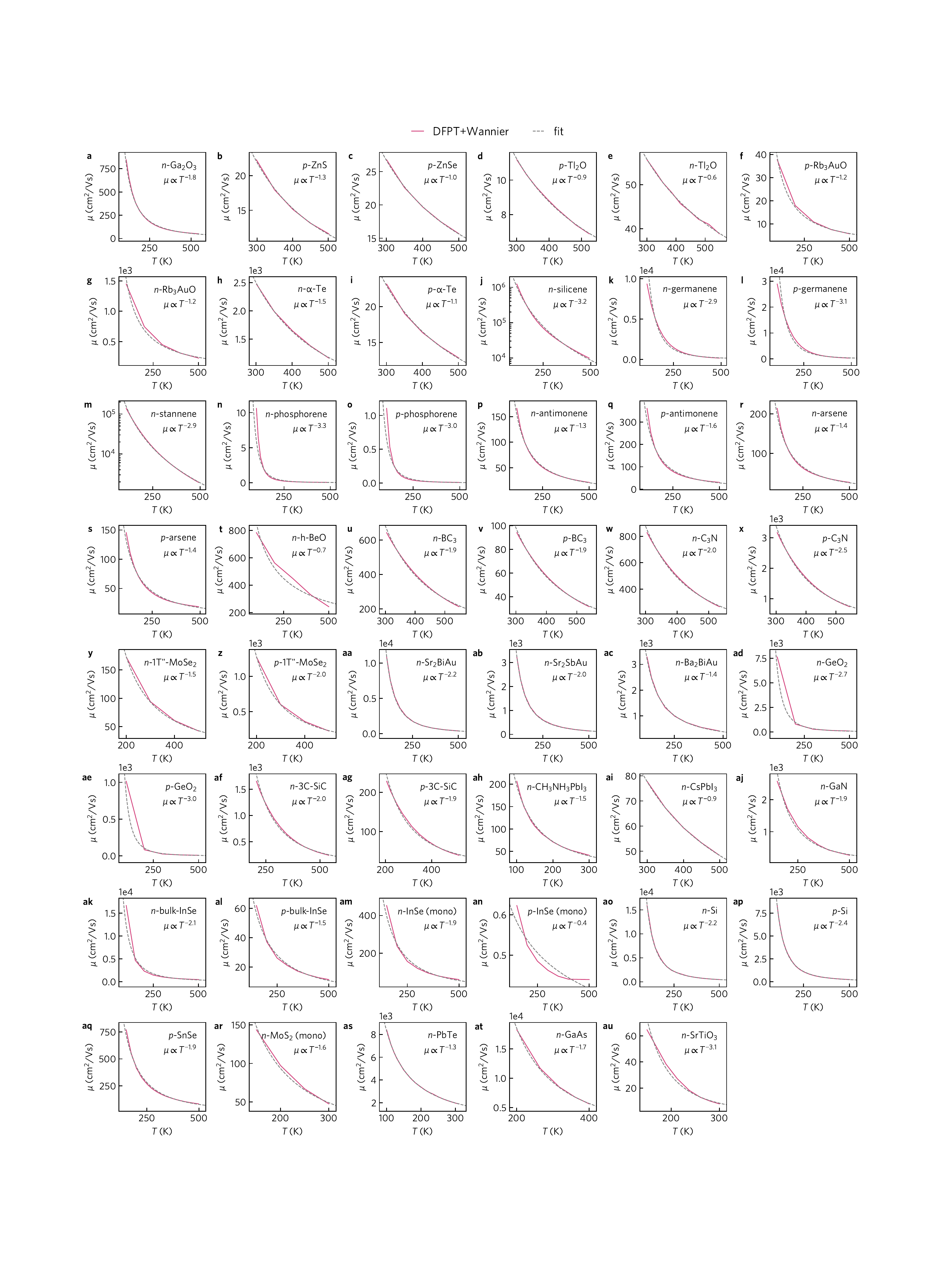}
}
    \vspace*{-3.5cm}
    \caption{Temperature against mobility (pink line) for all materials in the density functional perturbation theory combined with Wannier interpolation (DFPT+Wanner) dataset. The line of best fit, calculated from Eq.~\ref{eq:t-dependence-fit}, is shown in grey.} 
    \label{fig:dfpt-t-dep}
\end{figure}

In the main manuscript, we present a summary of the temperature dependence of mobility calculated from density functional perturbation theory combined with Wannier interpolation (DFPT+Wannier).
Our dataset includes 47 compounds and covers both $n$ and $p$-type doping, and bulk and monolayer structures.
The data was compiled from studies in which DFPT+Wannier calculations of mobility were performed at multiple temperatures.
In order to indicate the dominant scattering mechanism, we  only consider results that include a breakdown of different scattering types, such as the energy or \textbf{k}-dependence of the scattering rates separated into different mechanisms.
At the time of writing, our data set includes all DFPT+Wannier calculations that satisfy the above selection criteria.
Mobility data was extracted using WebPlotDigitizer \cite{Rohatgi2020}.
The temperature ($T$) dependence of mobility ($\mu$) was calculated by fitting the mobility to
\begin{equation}
    \mu (T) = T^{a} + b,
    \label{eq:t-dependence-fit}
\end{equation}
where $a$ and $b$ are fitted parameters.
The temperature dependence of mobility for each materials in the dataset are listed in Table.~\ref{tab:dfpt-all}.
Furthermore, in Fig.~\ref{fig:dfpt-t-dep}, we plot the full temperature dependent results and include the line of best fit calculated from Eq.~\ref{eq:t-dependence-fit}.

\begin{spacing}{1}
\begin{longtable}{lllrlrl}
\caption{Temperature dependence of mobility for all materials calculated by density functional perturbation theory combined with Wannier interpolation (DFPT+Wannier). We indicate whether the doping is $n$ or $p$-type, the dominant scattering mechanism, and the effective mass of majority carriers ($m^{*}$). The scattering mechanisms identified are deformation potential scattering (DP), polar optical phonon scattering (PO), piezoelectric scattering (PI), and polaronic effects (PL). The unit of effective mass is the electron rest mass ($m_0$).}
\label{tab:dfpt-all}\\
\toprule
                  Formula & Doping &  Topology &  $T$ dependence & Mechanism &  $m^{*}$ ($m_0$) &                                      Ref. \\
\midrule
\endfirsthead
\toprule
                  Formula & Doping &  Topology &  $T$ dependence & Mechanism &  $m^{*}$ ($m_0$) &                                      Ref. \\
\midrule
\endhead
\midrule
\endfoot

\bottomrule
\endlastfoot
              $\alpha$-Te &    $p$ & monolayer &           -1.15 &        DP &     0.24 &    \citenum{ma2020StrainInducedUltrahigh} \\
              $\alpha$-Te &    $n$ & monolayer &           -1.47 &        DP &     0.10 &    \citenum{ma2020StrainInducedUltrahigh} \\
           1T"-MoSe$_{2}$ &    $p$ & monolayer &           -1.98 &        DP &     0.54 &       \citenum{ge2020LargeThermoelectric} \\
           1T"-MoSe$_{2}$ &    $n$ & monolayer &           -1.52 &        DP &     0.94 &       \citenum{ge2020LargeThermoelectric} \\
                   3C-SiC &    $p$ &      bulk &           -1.90 &    PO, DP &     0.45 &    \citenum{meng2019PhononlimitedCarrier} \\
                   3C-SiC &    $n$ &      bulk &           -2.04 &    PO, DP &     0.31 &    \citenum{meng2019PhononlimitedCarrier} \\
                 BC$_{3}$ &    $p$ & monolayer &           -1.90 &        DP &     0.50 &     \citenum{su2020PhononlimitedMobility} \\
                 BC$_{3}$ &    $n$ & monolayer &           -1.89 &        DP &     0.30 &     \citenum{su2020PhononlimitedMobility} \\
             Ba$_{2}$BiAu &    $n$ &      bulk &           -1.42 &        PO &     0.16 &      \citenum{park2019HighThermoelectric} \\
                 C$_{3}$N &    $n$ & monolayer &           -2.03 &        DP &     0.45 &     \citenum{su2020PhononlimitedMobility} \\
                 C$_{3}$N &    $p$ & monolayer &           -2.50 &        DP &     0.24 &     \citenum{su2020PhononlimitedMobility} \\
CH$_{3}$NH$_{3}$PbI$_{3}$ &    $n$ &      bulk &           -1.49 &        PO &     0.21 &              \citenum{ponce2019OriginLow} \\
              CsPbI$_{3}$ &    $n$ &      bulk &           -0.93 &        PO &     0.17 &              \citenum{ponce2019OriginLow} \\
          Ga$_{2}$O$_{3}$ &    $n$ &      bulk &           -1.75 &    PO, DP &     0.25 &          \citenum{ma2020ElectronMobility} \\
                     GaAs &    $n$ &      bulk &           -1.70 &        PO &     0.06 &  \citenum{lee2020InitioElectrontwophonon} \\
                      GaN &    $n$ &      bulk &           -1.91 &    PO, PI &     0.20 &              \citenum{ponce2019OriginLow} \\
                GeO$_{2}$ &    $n$ &      bulk &           -2.73 &        PO &     0.29 &         \citenum{bushick2020ElectronHole} \\
                GeO$_{2}$ &    $p$ &      bulk &           -2.96 &        PO &     1.47 &         \citenum{bushick2020ElectronHole} \\
                     InSe &    $p$ & monolayer &           -0.41 &        DP &     3.40 &      \citenum{li2019DimensionalCrossover} \\
                     InSe &    $p$ &      bulk &           -1.55 &        PO &     2.60 &      \citenum{li2019DimensionalCrossover} \\
                     InSe &    $n$ & monolayer &           -1.86 &        PO &     0.16 &      \citenum{li2019DimensionalCrossover} \\
                     InSe &    $n$ &      bulk &           -2.13 &        PO &     0.10 &      \citenum{li2019DimensionalCrossover} \\
                MoS$_{2}$ &    $n$ & monolayer &           -1.59 &        PO &     0.49 &     \citenum{zhao2018IntrinsicElectronic} \\
                     PbTe &    $n$ &      bulk &           -1.30 &        PO &     0.05 &   \citenum{cao2018DominantElectronphonon} \\
              Rb$_{3}$AuO &    $n$ &      bulk &           -1.16 &        PO &     0.27 &     \citenum{zhao2020AnomalousElectronic} \\
              Rb$_{3}$AuO &    $p$ &      bulk &           -1.17 &        PO &     0.23 &     \citenum{zhao2020AnomalousElectronic} \\
                       Si &    $n$ &      bulk &           -2.20 &        DP &     0.26 &     \citenum{ponce2018PredictiveManybody} \\
                       Si &    $p$ &      bulk &           -2.37 &        DP &     0.22 &     \citenum{ponce2018PredictiveManybody} \\
                     SnSe &    $p$ &      bulk &           -1.86 &        PO &     0.22 &    \citenum{ma2018IntrinsicPhononlimited} \\
             Sr$_{2}$BiAu &    $n$ &      bulk &           -2.20 &    PO, DP &     0.21 &      \citenum{park2020HighThermoelectric} \\
             Sr$_{2}$SbAu &    $n$ &      bulk &           -2.05 &    PO, DP &     0.21 &      \citenum{park2020HighThermoelectric} \\
              SrTiO$_{3}$ &    $n$ &      bulk &           -3.09 &    PO, PL &     0.75 &        \citenum{zhou2019PredictingCharge} \\
                Tl$_{2}$O &    $n$ & monolayer &           -0.59 &        PO &     0.15 &     \citenum{huang2020HighThermoelectric} \\
                Tl$_{2}$O &    $p$ & monolayer &           -0.87 &    PO, DP &     0.28 &     \citenum{huang2020HighThermoelectric} \\
                      ZnS &    $p$ &      bulk &           -1.26 &    PO, DP &     0.55 & \citenum{ding2021ThermoelectricTransport} \\
                     ZnSe &    $p$ &      bulk &           -1.05 &    PO, DP &     0.75 & \citenum{ding2021ThermoelectricTransport} \\
               antimonene &    $n$ & monolayer &           -1.33 &        DP &     0.21 &       \citenum{wu2020AccuratePredictions} \\
               antimonene &    $p$ & monolayer &           -1.60 &        DP &     0.46 &       \citenum{wu2020AccuratePredictions} \\
                   arsene &    $p$ & monolayer &           -1.40 &        DP &     0.53 &       \citenum{wu2020AccuratePredictions} \\
                   arsene &    $n$ & monolayer &           -1.37 &        DP &     0.23 &       \citenum{wu2020AccuratePredictions} \\
                germanene &    $p$ & monolayer &           -3.14 &        DP &     0.02 &       \citenum{wu2020AccuratePredictions} \\
                germanene &    $n$ & monolayer &           -2.85 &        DP &     0.02 &       \citenum{wu2020AccuratePredictions} \\
                    h-BeO &    $n$ & monolayer &           -0.70 &        PO &     0.79 &   \citenum{ge2020PhononlimitedElectronic} \\
              phosphorene &    $p$ & monolayer &           -3.01 &        DP &     0.85 &       \citenum{wu2020AccuratePredictions} \\
              phosphorene &    $n$ & monolayer &           -3.32 &        DP &     0.44 &       \citenum{wu2020AccuratePredictions} \\
                 silicene &    $n$ & monolayer &           -3.18 &        DP &     0.01 &       \citenum{wu2020AccuratePredictions} \\
                 stannene &    $n$ & monolayer &           -2.93 &        DP &     0.03 &       \citenum{wu2020AccuratePredictions} \\
\end{longtable}

\end{spacing}

\clearpage
\section{Temperature dependence in a single parabolic band}

\begin{figure}
    \centering
    \includegraphics[width=0.94\textwidth]{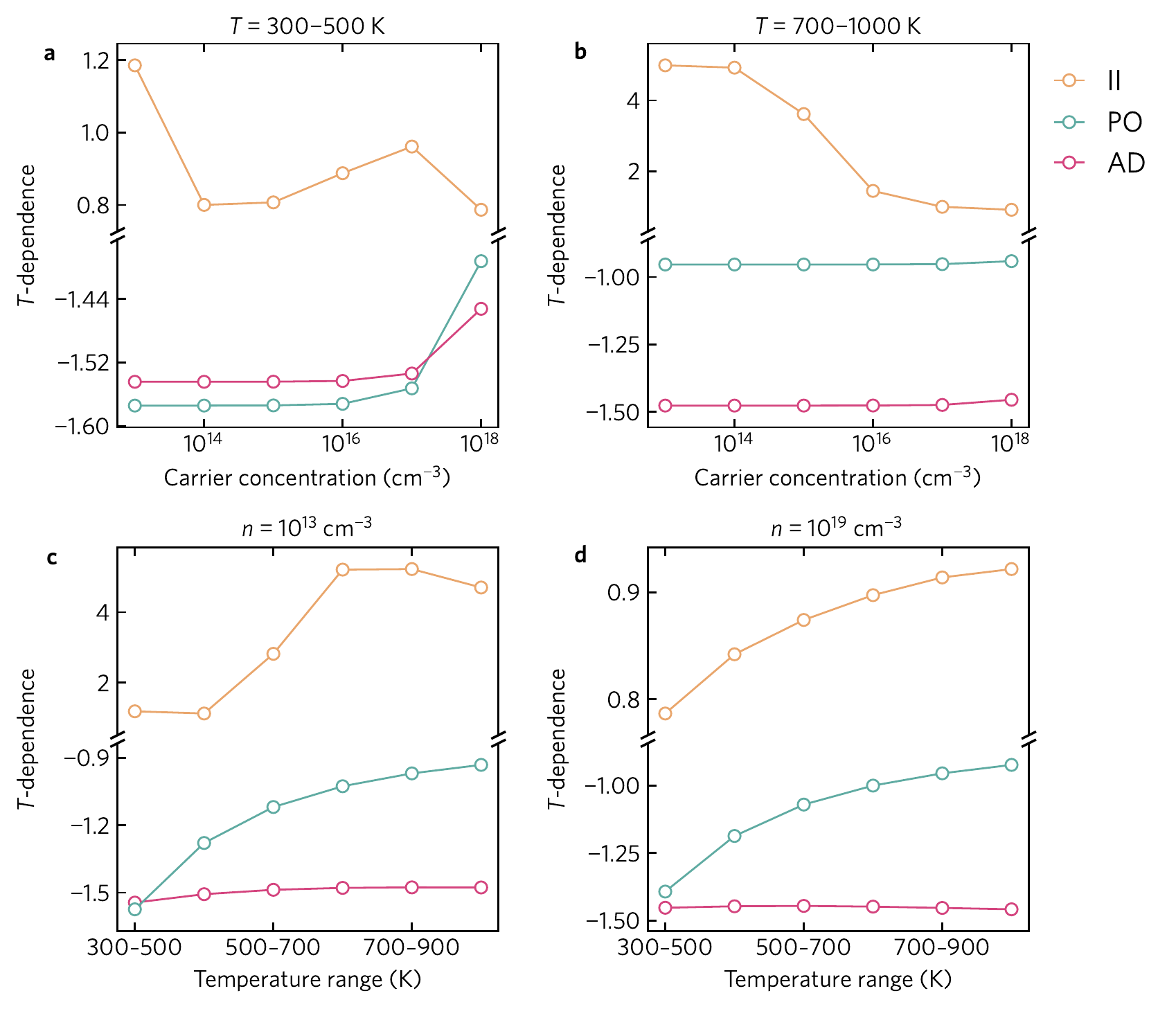}
    \caption{Increasing the temperature range or carrier concentration weakens (makes less negative) the temperature dependence of mobility. Calculations performed in a single parabolic band using polar optical phonon scattering (PO), acoustic deformation potential scattering (AD), and ionized impurity scattering (II). \textbf{a,b} The temperature dependence of mobility against carrier concentration at \textbf{a} low temperatures (\SIrange{300}{500}{\kelvin}) and \textbf{b} high temperatures (\SIrange{800}{1000}{\kelvin}). \textbf{c,d} The temperature dependence of mobility against the temperature range in which mobility is measured at \textbf{c} low carrier concentrations (\SI{e13}{\per\cubic\cm}) and \textbf{d} high carrier concentrations (\SI{e18}{\per\cubic\cm}).
    }
    \label{fig:si_conc_and_t_dep}
\end{figure}

\clearpage
\section{Impact of band structure on temperature dependence}

\begin{figure}
    \centering
    \includegraphics[width=0.85\textwidth]{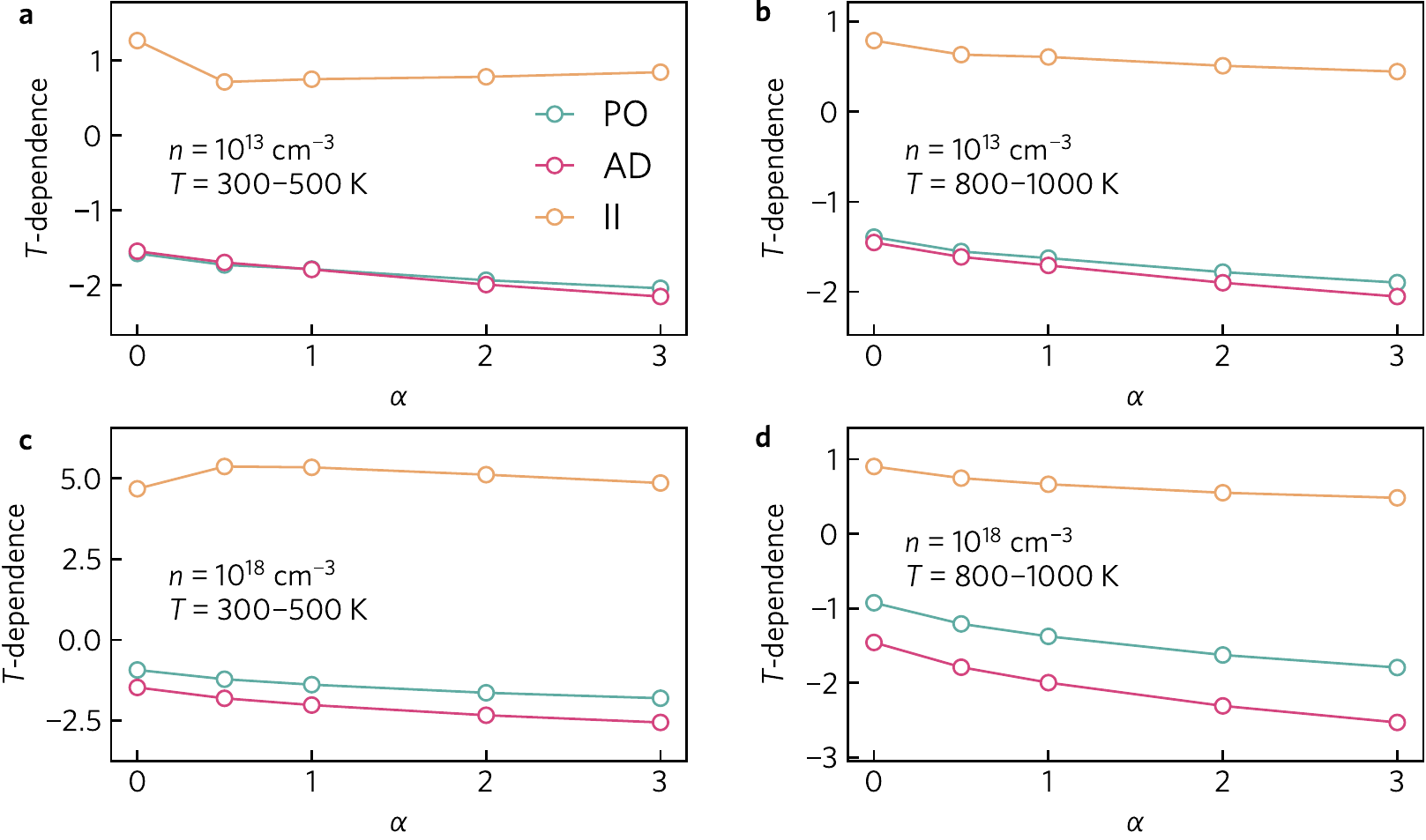}
    \caption{Effect of Kane non-parabolicity parameter ($\alpha$) on the temperature ($T$) dependence of mobility calculated using polar optical phonon scattering (PO), acoustic deformation potential scattering (AD), and ionized impurity scattering (II). We have considered all combinations of low (\SIrange{300}{500}{\kelvin}) and high (\SIrange{800}{1000}{\kelvin}) temperature regimes, and low (\SI{e13}{\per\cubic\centi\meter}) and high (\SI{e18}{\per\cubic\centi\meter}) carrier concentrations.}
    \label{fig:non-para}
\end{figure}

In the main text, we investigate the impact of non-isotropic and non-parabolic bands on the $T$ dependence of mobility.
To include non-parabolicity, we generated a series of band structures according to the Kane quasi-linear dispersion relation \cite{kane1957BandStructure}
\begin{equation}
    \frac{\hbar^2 k}{2 m^{*}} = \varepsilon (1 + \alpha\varepsilon),
    \label{eq:kane}
\end{equation} 
where $\hbar$ is the reduced Planck constant, $\varepsilon$ is the band energy, $k = \left | \mathbf{k} \right |$ is the distance in reciprocal space of \textbf{k} from the band edge (set to the $\Gamma$-point $[0, 0, 0]$), and $m^{*}$ is the effective mass of charge carriers. 
A larger value of $\alpha$ (Kane parameter) indicates increased linearity of the band away from the band minimum.
Due to periodicity, the eigenvalues of Bloch functions exhibit a derivative of zero at the Brillouin zone boundary.
To add this physical behaviour and ensure that the band structures could be smoothly Fourier interpolated (as required by \textsc{amset}) we added a bump function to the dispersion generated by Eq.~\ref{eq:kane} according to the procedure outlined in Ref.~\citenum{tu2011IntroductionManifolds}.
The bump function becomes active at \textbf{k}-points greater than \SI{80}{\percent} of the distance from the zone center to the zone boundary.
Even at the largest carrier concentrations used in our tests (\SI{e18}{\per\cubic\centi\meter}), the Fermi--Dirac occupation in the bump region is extremely small.
Accordingly, the transport results will be entirely dictated by the quasi-linear dispersion region around the band edge.
Calculations included polar optical phonon, acoustic deformation potential, and ionized impurity scattering and were performed using the \textsc{amset} package \cite{ganose2021EfficientCalculation}.
Due to lack of wave function information, the wave function overlaps in the scattering matrix elements were set to unity (indicting complete overlap).
This is a reasonable approximation for a single near-parabolic band and is consistent with the approximations made in the simplified models of transport historically used to determine the temperature dependence of mobility.
We considered both low (\SIrange{300}{500}{\kelvin}) and high (\SIrange{800}{1000}{\kelvin}) temperature regimes, and low (\SI{e13}{\per\cubic\centi\meter}) and high (\SI{e18}{\per\cubic\centi\meter}) carrier concentrations.
The full results for all temperature and doping combinations are presented in Fig.~\ref{fig:non-para}.

\begin{figure}
    \centering
    \includegraphics[width=0.85\textwidth]{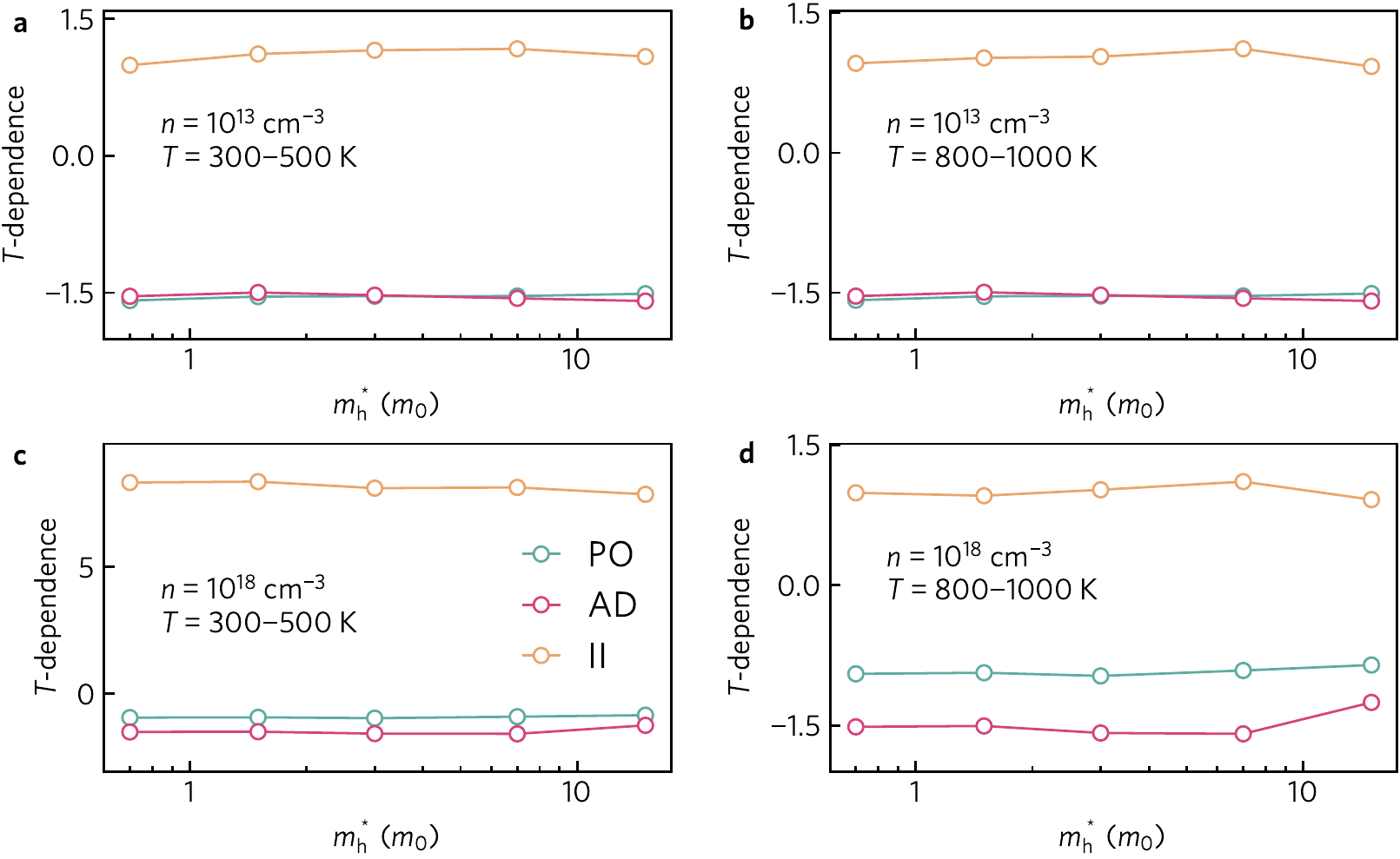}
    \caption{Effect of a heavy anisotropic effective mass ($m_\mathrm{h}^{*}$) on the temperature ($T$) dependence of mobility calculated using polar optical phonon scattering (PO), acoustic deformation potential scattering (AD), and ionized impurity scattering (II). We have considered all combinations of low (\SIrange{300}{500}{\kelvin}) and high (\SIrange{800}{1000}{\kelvin}) temperature regimes, and low (\SI{e13}{\per\cubic\centi\meter}) and high (\SI{e18}{\per\cubic\centi\meter}) carrier concentrations.}
    \label{fig:aniso}
\end{figure}

We have also investigated the impact of band anisotropy by constructing a series of band structures with varying effective masses in each principal direction.
We enforce a simple cubic symmetry in which the effective masses along $\Gamma$--M and $\Gamma$--R are held the same while the mass along $\Gamma$--X is increased from \SI{0.7}{\electronmass} to \SI{30}{\electronmass}.
These anisotropic band structures are generated by smoothly inverting upward paraboloids into downward paraboloids after the methods of \cite{Park2021OptimalBandStructure}, whereby the derivative continuity is enforced at the point of inflection. By doing so, we ensure that the entire band structure is parabolic while still bounded by the Brillouin zone, crossing the zone boundary orthogonally. We also set the inflection point to be the halfway point to the zone boundary in each direction. This ensures that the band structure retains the same directional effective masses after inversion, save for the sign, such that the entire band structure is described by three directional effective masses.
Scattering calculations included polar optical phonon, acoustic deformation potential, and ionized impurity scattering and were performed using the \textsc{amset} package \cite{ganose2021EfficientCalculation}.
Since such arbitrary band structures lack wave function information, the wave function overlaps in the scattering matrix elements were set to unity (indicating complete overlap).
We considered both low (\SIrange{300}{500}{\kelvin}) and high (\SIrange{800}{1000}{\kelvin}) temperature regimes, and low (\SI{e13}{\per\cubic\centi\meter}) and high (\SI{e18}{\per\cubic\centi\meter}) carrier concentrations.
The full results for all temperature and doping combinations are presented in Fig.~\ref{fig:aniso}

\clearpage
\section{Additional impacts of phonon frequency on the temperature dependence of mobility}


The temperature dependence of mobility is also compounded by the fraction of occupied electronic states above the phonon emission threshold (Fig.~3b in the main text).
We calculate this fraction as $\int_{\varepsilon_\mathrm{po}}^{\infty} [ D(\varepsilon)\pdv{f}{\varepsilon}]d\varepsilon / \int_0^{\infty} [ D(\varepsilon)\pdv{f}{\varepsilon}]d\varepsilon$,  where $D$ is the density of states, $\pdv{f}{\varepsilon}$ is the derivative of the Fermi--Dirac distribution and $\varepsilon_\mathrm{po} = 2\pi \hbar \omega_\mathrm{po}$ is the energy of the polar optical phonon mode.
As revealed in Fig.~3e in the main text, at \SI{300}{\kelvin} and at low phonon frequencies (\SI{4}{\tera\hertz}) over \SI{80}{\percent} of the states that contribute to the mobility originate from above the phonon emission threshold.
As the temperature increases to \SI{500}{\kelvin}, the fraction of post-emission states only increases by \SI{10}{\percent} (a relative increase of \SI{12.5}{\percent}).
In contrast, when the phonon frequency is large (\SI{20}{\tera\hertz}), the post-emission states account for only \SI{10}{\percent} of the states that contribute to the mobility at \SI{300}{\kelvin}.
However, as the temperature increases, the fraction of post-emission states increases rapidly to $\sim$\SI{30}{\percent} at \SI{500}{\kelvin}, a relative increase of \SI{200}{\percent}.
This is critical because the lifetimes of the post-emission states near the emission onset are approximately an order of magnitude shorter than those of the pre-emission states.
Accordingly, this dramatic increase in the contribution of post emissions states results in a considerable decrease in the electron mobility at high temperatures, leading to a quicker decay of mobility with temperature.

At higher, ``post-Debye'' temperatures (\SIrange{800}{1000}{\kelvin}), phonon occupation essentially increases linearly as $n\propto T$, resulting in more gradual decay of electron lifetimes with temperature (Fig.~3c,d in the main text).
Furthermore, at high temperatures, the post-emission states account for over \SI{50}{\percent} of all states that contribute to mobility across all phonon frequencies (Fig.~3e in the main text).
Together these combine to drastically reduce the temperature dependence of mobility, particularly for large $\omega_\mathrm{po}$ (Fig.~3a in the main text).
For example, at a phonon frequency of \SI{20}{\tera\hertz} the mobility dependence is reduced from $T^{-3.34}$ between \SIrange{300}{500}{\kelvin} to $T^{-1.81}$ between \SIrange{800}{1000}{\kelvin}. 

Thermoelectric materials generally exhibit optimal performance at ``high doping''.
As demonstrated in Fig.~3a in the main text, an increase in the electron concentration from \SI{e13}{\per\cubic\centi\meter} to \SI{e18}{\per\cubic\centi\meter} results in a weakening of the temperature dependence across all phonon frequencies.
This behaviour can again be rationalized by considering the fraction of post-emission states that contribute to the mobility.
For phonon frequencies below $\sim$\SI{12}{\tera\hertz} almost all the occupied electronic states ($>$\SI{99}{\percent}) are post-emission (Fig.~3f in the main text).
The lifetimes of post-emission states exhibit a relatively weak temperature dependence (between $\tau \propto T^{-0.98}$--$T^{-0.81}$, see Fig.~3d in the main text) and result in a mobility trend that is always less negative than $\mu \propto T^{-1.3}$.
At larger phonon frequencies, the fraction of post-emission states decreases to as low as \SI{30}{\percent} at \SI{300}{\kelvin}, and the mobility picks up additional temperature dependence due to the potential for thermal activation of these states.

\clearpage
\section{Generation of AMSET dataset}

\begin{figure}
    \centering
    \includegraphics[width=0.5\linewidth]{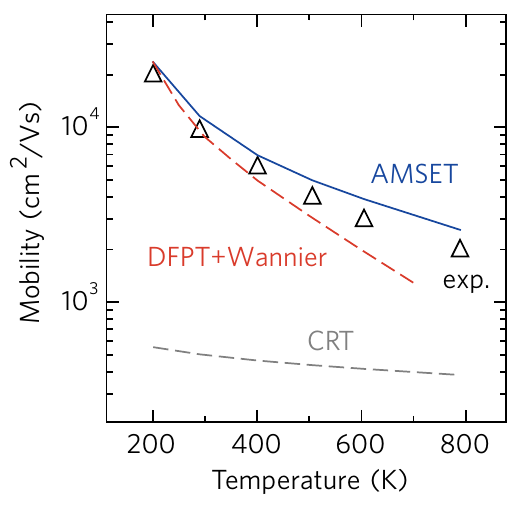}
    \caption{\textsc{amset} reproduces the experimental mobility of $n$-GaAs. Temperature against mobility for $n$-GaAs at low doping (\SI{e13}{\per\cubic\centi\meter}). Experimental measurements (black triangles, Ref.~\citenum{rode1971ElectronTransporta}) compared against \textsc{amset} (blue line), density functional perturbation theory combined with Wannier interpolation (DFPT+Wannier, red line, Ref.~\citenum{zhou2016InitioElectron}) and the constant relaxation time approximation (grey line) with the lifetime set to \SI{e-14}{\second}.}
    \label{fig:amset-gaas}
\end{figure}

In the main manuscript, we present the results of 23,000 mobility calculations performed using the \textsc{amset} package.
\textsc{amset} enables the explicit calculation of charge carrier scattering rates and the solution of the Boltzmann transport equation to obtain transport properties.
\textsc{amset} has been shown provide scattering rates and mobility within $\sim$ \SI{10}{\percent} of DFPT+Wannier when benchmarked on 23 semiconductors \cite{ganose2021EfficientCalculation}.
This is highlighted in Fig.~\ref{fig:amset-gaas}, where we demonstrate that \textsc{amset} can reproduce the experimental mobility of GaAs across the full temperature range.

The main inputs for \textsc{amset} are an electronic band structure and common materials parameters, including the static and high-frequency dielectric constants, elastic constant, polar phonon frequency, and deformation potentials.
There are four key differences between the \textsc{amset} calculations performed in this work and the calculations in Ref.~\citenum{ganose2021EfficientCalculation}.
Firstly, the wave function overlaps that enter the scattering matrix elements can be computed using the first principles wave functions directly or approximated using the orbital projections, as outlined in \citet{rode1975ChapterLowField}.
As the wave functions were not available for the 23,000 materials used in this dataset, we have used the latter approach when calculating the matrix elements.
Secondly, in Ref.~\citenum{ganose2021EfficientCalculation}, the materials parameters were all obtained from first principles calculations.
In this work, we have used a combination of first principles calculations and machine learning to obtain the materials parameters for each material.
Thirdly, in this work we use only two values for the deformation potential (one for the valence bands and one for the conduction bands) which are constant across all \textbf{k}-points, whereas in Ref.~\citenum{ganose2021EfficientCalculation} the deformation potential is a tensor that depends on the band index and \textbf{k}-point.
Finally, \textsc{amset} has the capacity to exploit any anisotropy in material parameters by using full property tensors.
However, due to the difficulty of machine learning a full tensor, in this work we used isotropically averaged materials parameters if the first principles result was not available.
While these approximations will result in a loss of accuracy, we stress that our goal is only to simulate a realistic distribution of transport properties rather than predict exact values for specific materials.
Furthermore, while some anisotropy will be lost by the averaging of tensor properties, we still account for any anisotropy in the electronic band structure by use of the full \textbf{k}-dependent group velocities and eigenvalues.

\textsc{amset} calculations were performed on uniform electronic band structures taken from the Materials Project.\cite{jainCommentaryMaterialsProject2013}
Band structures were calculated using density functional theory (DFT) with the Perdew--Burke--Ernzerhof\cite{perdewGeneralizedGradientApproximation1996} (PBE) exchange correlation functional as implemented in the Vienna \textit{ab initio} simulation package (VASP).\cite{Kresse1996,Kresse1996a}
Corrections for the Hubbard U were implemented for cobalt, chromium, iron, manganese, molybdenum, nickel, vanadium, and tungsten in oxides using the standard Materials Project values.\cite{jain2011FormationEnthalpies}
Calculations were performed in a plane-wave basis set with scalar relativistic psueodpoentials and with the interactions between core and valence electrons described using the projector augmented-wave method (PAW).\cite{blochlProjectorAugmentedwaveMethod1994,Kresse1999}
The plane-wave energy cutoff was set to 520 eV.
Several versions of the pymatgen\cite{ong2013PythonMaterials} input sets were been used to calculate the band structures.
The primary difference of between the inputs --- in the context of \textsc{amset} transport calculations --- is the density of the \textbf{k}-point mesh given in number of \textbf{k}-points per reciprocal volume.
The majority of calculations in this dataset (\SI{80}{\percent}) used a \textbf{k}-point density of \SI{100}{\textbf{k}{\mhyphen}points / \angstrom^{-3}}, with the remainder using densities of \SI{500}{\textbf{k}{\mhyphen}points / \angstrom^{-3}} (\SI{8}{\percent}), \SI{1000}{\textbf{k}{\mhyphen}points / \angstrom^{-3}} (\SI{8}{\percent}), and \SI{2000}{\textbf{k}{\mhyphen}points / \angstrom^{-3}} (\SI{4}{\percent}).
The band structures were accessed using the Materials Project application programming interface (API) available in pymatgen.\cite{ong2013PythonMaterials}
The corresponding wave functions were not available, so we have used the orbital projections at each band and k-point to approximate the wave function overlap.
We have considered all materials in the Materials Project database for which the band gap is greater than \SI{0.05}{\electronvolt} and for which there is a uniform band structure available.

\subsection{Materials parameter datasets}

To obtain the materials parameters needed to compute scattering rates for all 23,000 materials, we have employed a combination of first principles databases and machine learning.
In this section, we outline the generation and cleaning of the datasets.
The Materials Project database contains 13,421 elastic constant tensors and 8,570 dielectric tensors (and Born effective charges and $\Gamma$-point phonon frequencies as these are generated in the same calculation).\cite{jainCommentaryMaterialsProject2013,dejong2015ChartingComplete,petousis2017HighthroughputScreening}
We used the Born effective charges and phonon frequencies to calculate the polar phonon frequency, needed for polar optical phonon scattering, according to the methodology outlined in Ref.~\cite{ganose2021EfficientCalculation}.
In addition, we generated a dataset of 995 valence and conduction band acoustic deformation potentials.
The 995 materials were a subset of the semiconductor materials for which elastic constant tensors were available. 
The deformation potentials were calculated similarly to the methodology outlined in Ref.~\citenum{ganose2021EfficientCalculation}, however, we only consider two deformation potentials --- one for the valence band maximum and one for the conduction band minimum --- which are constant across all bands and \textbf{k}-points.
Finally, we used the dataset of 6,030 band gaps calculated using the HSE06 exchange--correlation functional prepared by \citet{jie2019NewMaterialGo}.
In Table \ref{tab:ml-datasets}, we summarise the source and number of materials in each dataset.

\begin{table}
\begin{tabular}{@{}llllll@{}}
\toprule
Property             & $N$ materials (raw) & $N$ materials (cleaned) & RMSE                     & $r^2$ & Ref.                                          \\ \midrule
$E_\mathrm{g}^\mathrm{HSE06}$ & 6,030 & 5,658$^\mathrm{a}$ & \SI{0.97}{\electronvolt} & 0.78 & \citenum{jie2019NewMaterialGo}                \\
$C$                  & 13,421            & 11,160$^\mathrm{a,b,c}$              & \SI{26.1}{\giga\pascal}  & 0.92  & \citenum{dejong2015ChartingComplete}          \\
$D_\mathrm{vbm}$     & 995               & 954$^\mathrm{a,d}$                 & \SI{1.49}{\electronvolt} & 0.40  & This work                                     \\
$D_\mathrm{cbm}$     & 995               & 954$^\mathrm{a,d}$                  & \SI{2.01}{\electronvolt} & 0.31  & This work                                     \\
$\epsilon_\mathrm{s}$         & 8,570 & 6,507$^\mathrm{a,e,f,g,h}$ & 3.41                     & 0.66 & \citenum{petousis2017HighthroughputScreening} \\
$\epsilon_\infty$    & 8,570             & 6,507$^\mathrm{a,e,f,g,h}$               & 1.24                     & 0.80  & \citenum{petousis2017HighthroughputScreening} \\
$\omega_\mathrm{po}$ & 8,569             & 6,507$^\mathrm{a,e,f,g,h}$              & \SI{1.68}{\tera\hertz}   & 0.88  & This work                                     \\ \bottomrule
\caption{Summary of datasets used for machine learning of materials properties. Properties include the band gap calculated using the HSE06 exchange--correlation functional ($E_\mathrm{g}^\mathrm{HSE06}$), elastic constant ($C$), acoustic deformation potential at the valence band maximum ($D_\mathrm{vbm}$) and conduction band minimum ($D_\mathrm{cbm}$), static ($\epsilon_\mathrm{s}$) and high-frequency ($\epsilon_\infty$) dielectric constants, and the polar optical phonon frequency ($\omega_\mathrm{po}$). The root mean squared error (RMSE) and $r^2$ correlation coefficient from 5-fold cross validation for the trained machine learning models are also provided\label{tab:ml-datasets}}
\end{tabular}
\footnotesize
\\
\vspace{-18pt}
\begin{flushleft}
$^\mathrm{a}$ Removed materials with a formation energy or energy above the convex hull more than 500 meV. \\
$^\mathrm{b}$ Removed materials with G$_\mathrm{Voigt}$, G$_\mathrm{Reuss}$, G$_\mathrm{VRH}$, K$_\mathrm{Voigt}$, K$_\mathrm{Reuss}$, or K$_\mathrm{VRH} < 1$. \\
$^\mathrm{c}$ Removed materials failing $\mathrm{G}_\mathrm{Voigt} \leq \mathrm{G}_\mathrm{VRH} \leq \mathrm{G}_\mathrm{Reuss}$ or $\mathrm{K}_\mathrm{Voigt} \leq \mathrm{K}_\mathrm{VRH} \leq \mathrm{K}_\mathrm{Reuss}$.\\
$^\mathrm{d}$ Removed materials with $D_\mathrm{vbm}$ or $D_\mathrm{cbm}$ $>$ \SI{30}{\electronvolt}.\\
$^\mathrm{e}$ Removed materials with a Materials Project band gap less than \SI{0.1}{\electronvolt}.\\
$^\mathrm{f}$ Removed materials with any imaginary phonon frequencies less (more negative) than \SI{-1}{\tera\hertz}.\\
$^\mathrm{g}$ Removed materials failing $\SI{300}{\tera\hertz} \geq \epsilon_\mathrm{s} \geq \epsilon_\infty \geq \SI{1}{\tera\hertz}$.\\
$^\mathrm{h}$ Removed materials with $\omega_\mathrm{po} < 0$.
\end{flushleft}
\end{table}

Before performing machine learning, each dataset was cleaned to remove unphysical materials properties.
Firstly, due to the complexity of machine learning tensors, we take the tensor average of the dielectric and elastic tensors.
Furthermore, for all datasets, we only keep materials which have a formation energy and energy above the convex hull less than \SI{500}{\milli\electronvolt} (as taken from the Materials Project).
Additional cleaning steps for each dataset are detailed in the footnotes of Table \ref{tab:ml-datasets} along with the  number of materials before and after data cleaning. 

\subsection{Machine learning of materials parameters}
\label{sec:ml}

Machine learning was performed using the \textsc{automatminer} package.\cite{dunn2020BenchmarkingMaterials}
\textsc{automatminer} is a tool for automatically creating machine learning pipelines for materials science datasets and includes automatic featurization with \textsc{matminer},\cite{ward2018MatminerOpen} feature reduction, and an automatic machine learning (AutoML) backend.
\textsc{automatminer} has been benchmarked on a diverse set of materials science challenges and shows comparable or better performance than bespoke hand tuned models.\cite{dunn2020BenchmarkingMaterials}
All models in this work included crystal structures as input and employed the ``express`` \textsc{automatminer} preset.
In addition to the features generated by \textsc{automatminer}, we also included the band gap calculated using PBE (with a Hubbard U correction if required) taken from the Materials Project.\cite{jainCommentaryMaterialsProject2013}

\begin{figure}
    \centering
    \includegraphics[width=\textwidth]{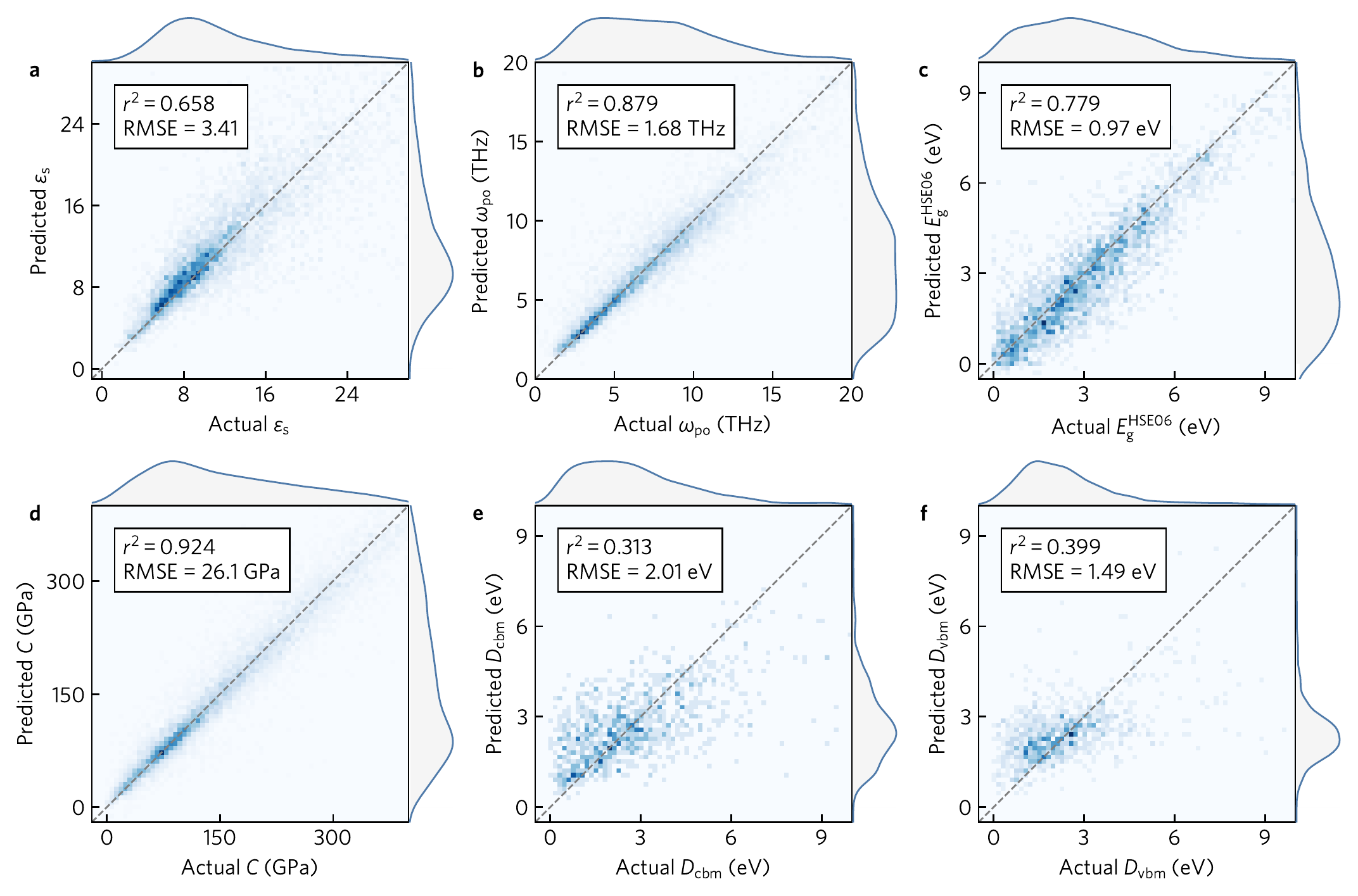}
    \caption{Cross validation predictions for machine learning models trained on the materials parameter datasets. \textbf{a} Static dielectric constant  ($\epsilon_s$), \textbf{b} polar phonon frequency ($\omega_\mathrm{po}$), \textbf{c} HSE06 band gap ($E_g^\mathrm{HSE06}$), \textbf{d} elastic constant ($C$), and \textbf{e}  acoustic deformation potentials at the valence band maximum ($D_\mathrm{vbm}$) and \textbf{f} conduction band minimum ($D_\mathrm{cbm}$). The root mean squared error (RMSE) of predictions and $r^2$ coefficient are given for each model.}
    \label{fig:ml-bechmark}
\end{figure}

We evaluated the models using 5-fold cross validation with the results presented in Fig.~\ref{fig:ml-bechmark}.
The mean root mean squared error (RMSE) of predictions and $r^2$ correlation coefficient for the 5-fold cross validation are also summarised in Table \ref{tab:ml-datasets}.
Overall, the models trained on larger datasets with more than 5,000 materials (elastic, dielectric, phonon frequency and HSE06 gap) are considerably more accurate, with $r^2$ values between 0.66--0.92.
The acoustic deformation potential datasets --- containing only 954 materials --- appear significantly more difficult, with $r^2$ values of 0.31 and 0.40 for the conduction band and valence band potentials, respectively.
Despite this, we note that the deformation potential is temperature independent and therefore will not affect the temperature-dependence of mobility and, thus, the conclusions in the main text.
Furthermore, our goal is only to simulate a realistic distribution of parameters rather than the exact parameter for each material, so strict accuracy is not essential.

\subsection{High-throughput AMSET workflow}

The machine learning models described in Sec.~\ref{sec:ml} were used to predict materials parameters for all 131,653 semiconductors in the Materials Project database.
\textsc{amset} calculations were performed on all materials that satisfied the following criteria:
(i) A uniform band structure with orbital projections was available;
(ii) The Materials Project band gap was greater than \SI{0.05}{\electronvolt};
(iii) The (predicted) HSE06 band gap was greater than \SI{0.05}{\electronvolt};
(iv) The (predicted) static dielectric constant ($\epsilon_\mathrm{s}$) was greater than the (predicted) high-frequency dielectric constant ($\epsilon_\infty$);
(v) All predicted materials properties were greater than zero.
In total this amounted to 23,572 materials.
We preferentially used materials properties calculated using first principles methods if they were available and satisfied the criteria outlined in the footnotes of Table \ref{tab:ml-datasets}.
In the case a first principles property was not available, we used the materials properties predicted from the machine learning models.
651 materials had all properties calculated from first principles, the remainder used a mixture of machine learned and first principles properties.
Finally, we applied a scissor correction to shift the band gap of the Materials Project band structure to match that of the (predicted) HSE06 gap.

\textsc{amset} calculations were performed in a high-throughput mode using the \textsc{fireworks} workflow management software.\cite{jain2015fireworks}
\textsc{amset} version \verb|0.4.3| was used throughout.
We used the default settings with the addition of free carrier screening of the polar optical phonon matrix element enabled by the \verb|free_carrier_screening| option.
Calculations were performed at both $n-$ and $p$-type doping concentrations and at temperatures ranging from \SIrange{200}{1000}{\kelvin}.
To ensure convergence of transport properties, the workflow performed several \textsc{amset} calculations with increasing \verb|interpolation_factor| --- the primary setting that controls the density of the interpolated \textbf{k}-point mesh --- until the change in mobility at all doping and temperatures was less than \SI{5}{\percent}.

\clearpage
\section{High doping AMSET screening}

\begin{figure}
    \centering
    \includegraphics[width=0.9\linewidth]{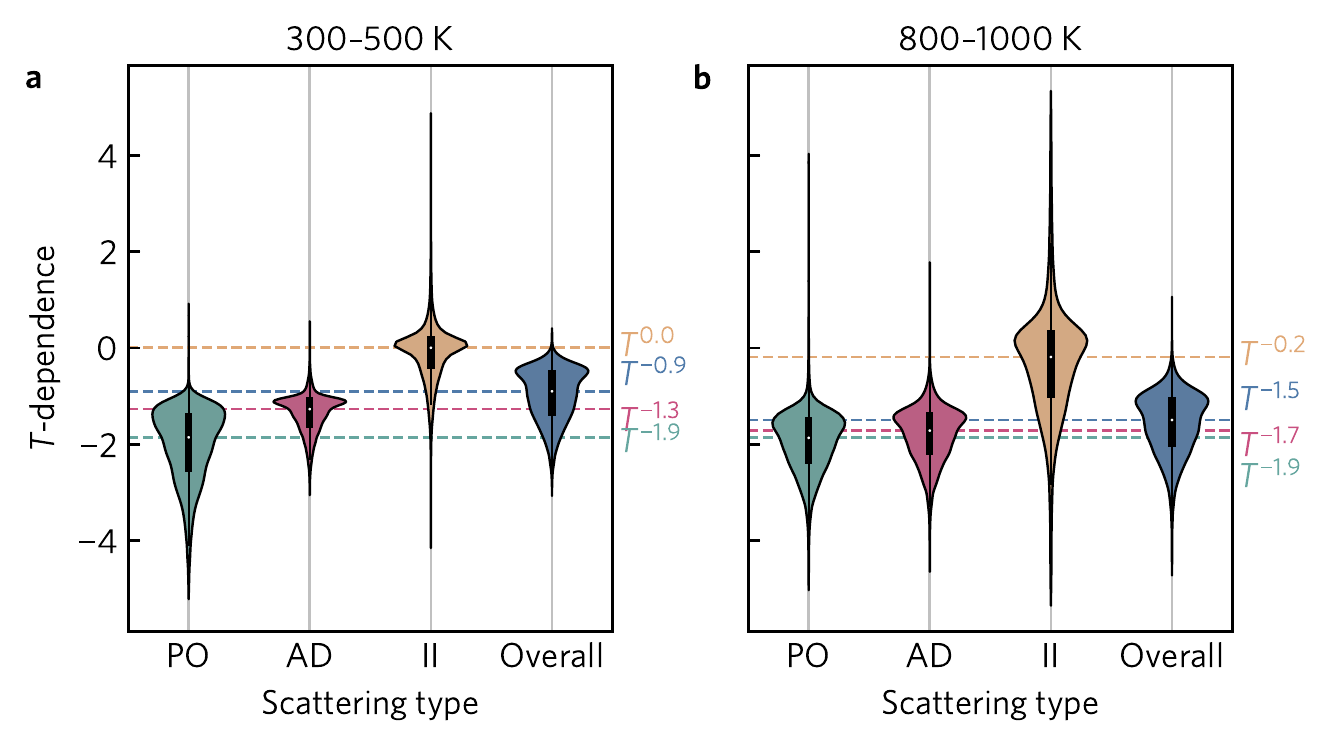}
    \caption{\textsc{amset} screening results at high doping (\SI{e19}{\per\cubic\centi\meter}). Histograms of the temperature ($T$) dependence of mobility for different scattering mechanisms at \textbf{a} low (\SIrange{300}{500}{\kelvin}) and \textbf{b} high (\SIrange{800}{1000}{\kelvin}) temperatures obtained from 23,000 materials. Calculations were performed using the \textsc{amset} package and are separated into polar optical (PO), acoustic deformation potential (AD), and ionized impurity (II) scattering. The ``Overall'' distribution gives the temperature dependence using the total scattering rate ($\tau^{-1}$) obtained from Matthiessen's rule ($\tau^{-1}_\mathrm{overall} = \tau^{-1}_\mathrm{po} + \tau^{-1}_\mathrm{ad} + \tau^{-1}_\mathrm{ii}$).}
\end{figure}

\clearpage
\section{Explaining the difference between the single parabolic band model and the AMSET distribution}

In the main text, we note that the temperature dependence resulting from the single parabolic band model acts as the upper bound to the AMSET distribution. We believe this originates from the tendency of bands in realistic materials to be less dispersive than a parabolic band ($E\propto k^{<2}$) rather than more dispersive ($E\propto k^{>2}$). I.e., the dispersion typically becomes linear away from the band edges. If $k^{<2}$, then group velocity increases more slowly with energy than for a parabolic band. This makes higher-energy carriers that get excited at higher temperatures less mobile, leading to a more negative temperature dependence. In contrast if $k^{>2}$, then higher-energy carriers would be relatively more mobile, leading to a less negative temperature dependence.

 
\clearpage
\section{Explaining the variation in temperature dependence at a particular polar phonon frequency}

\begin{figure}
    \centering
    \includegraphics[width=\textwidth]{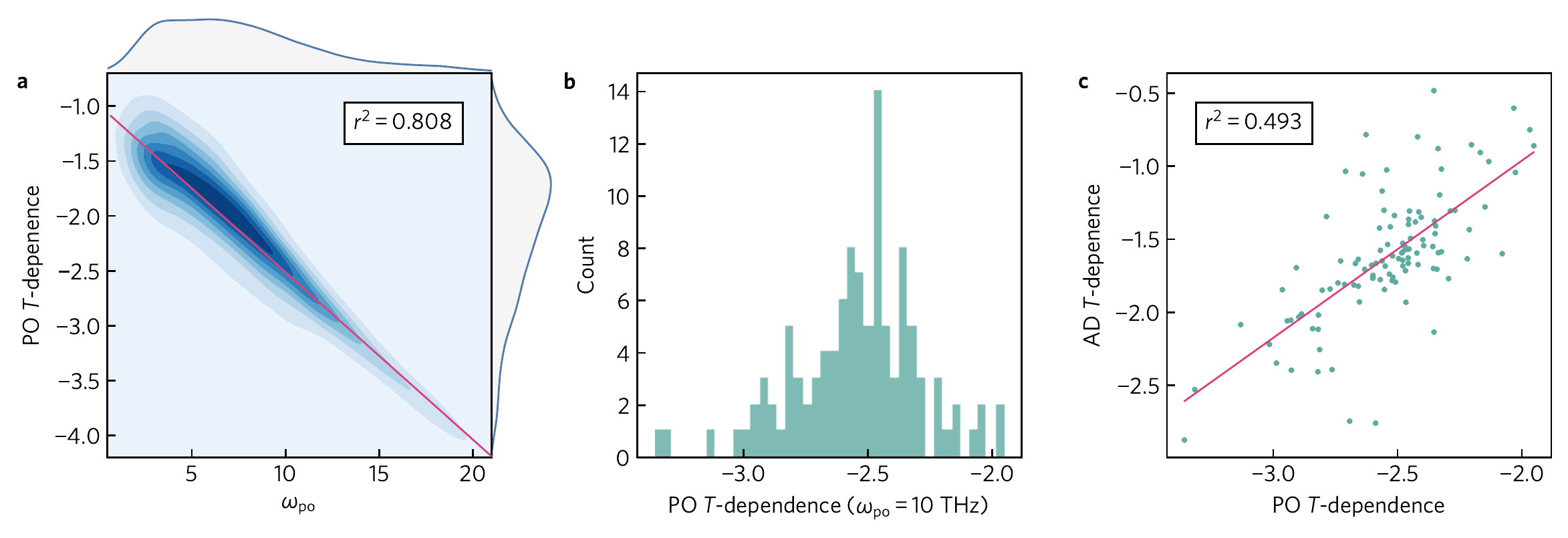}
    \caption{\textbf{a} Heatmap indicating the correlation between the polar optical (PO) phonon frequency ($\omega_\mathrm{po}$) and the temperature dependence of mobility (in the range \SIrange{300}{500}{\kelvin}) limited by optical phonon scattering. The $r^2 = 0.808$ coefficient indicates strong correlation. \textbf{b} Histogram of temperature dependence of mobility limited by optical phonon scattering for systems where $9.99 < \omega_\mathrm{po} < 10.01$. \textbf{c} The correlation between mobility limited by polar optical phonon scattering and acoustic deformation potential (AD) scattering. The $r^2 = 0.493$ coefficient indicates moderate correlation. In \textbf{a} and \textbf{c}, the pink line is the line of best fit, from which the $r^2$ value is calculated.}
    \label{fig:pop-t-dependence}
\end{figure}

As noted in the main text, the temperature dependence of mobility when limited by polar optical phonon scattering exhibits a strong correlation ($r^2 = 0.82$) to the polar phonon frequency, $\omega_\mathrm{po}$ (Fig.~\ref{fig:pop-t-dependence}a).
However, despite this correlation, the distribution of $T$ dependencies at any particular value of $\omega_\mathrm{po}$ is rather broad, indicating that $\omega_\mathrm{po}$ alone can not explain a materials $T$ dependence.
This is illustrated in Fig.~\ref{fig:pop-t-dependence}b, in which the $T$ dependencies of all materials with  $\SI{9.99}{\tera\hertz}<\omega_\mathrm{po}<\SI{10.01}{\tera\hertz}$ are depicted as a histogram.
While the distribution is centered at value of $\sim -2.5$, the tails of the distribution cover a wide range from $-3.36$ to $-1.95$.
Other factors responsible for the range of $T$ dependencies can include the shape of the bands (group velocities) and the behaviour of the scattering rates across the band structure.
We note that the other materials parameters that enter the polar optical phonon scattering matrix element, namely the static and high-frequency dielectric constants, are constant across all temperatures and therefore play no role in controlling the temperature dependence of mobility.
The behaviour of scattering rates across the band structure is computationally expensive to obtain and therefore cannot be used as an efficient predictor of the $T$-dependence.
Furthermore, as demonstrated by the correlation plot in Fig.~\ref{fig:pop-t-dependence}c, for materials with $\SI{9.99}{\tera\hertz}<\omega_\mathrm{po}<\SI{10.01}{\tera\hertz}$ there is a moderate correlation ($r^2 = 0.49$) between the temperature dependence of mobility limited by polar optical scattering and acoustic deformation potential scattering.
This correlation suggests that band structure features are at least partially responsible for the distribution of temperature dependencies at a specific $\omega_\mathrm{po}$.
In this section, we explore whether common band structure features can be used to predict the $T$ dependence of mobility.

\begin{figure}
    \centering
    \includegraphics[width=\textwidth]{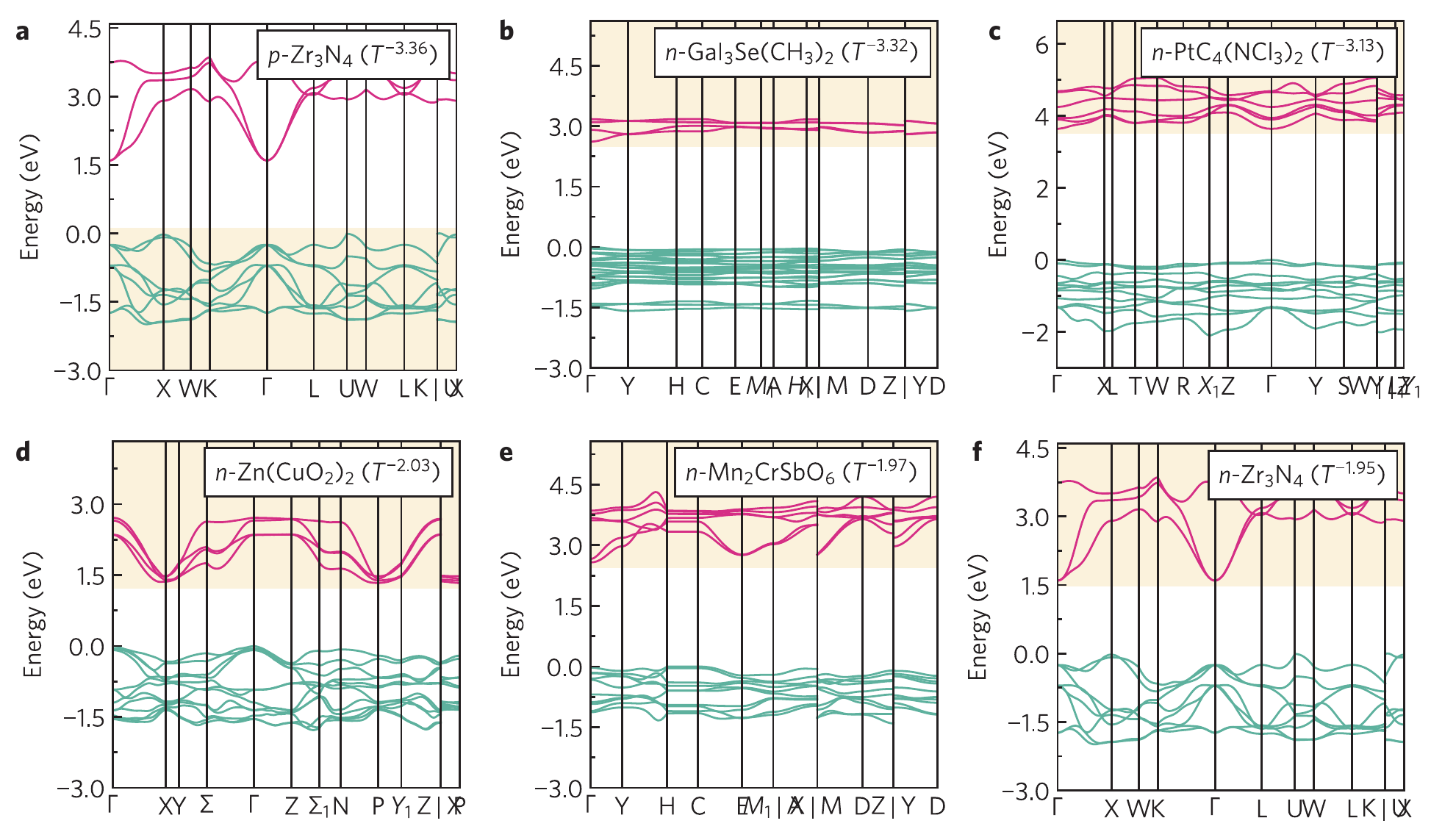}
    \caption{Band structures of the materials with the largest and smallest temperature dependence of mobility limited by polar optical phonon scattering, where $9.99 < \omega_\mathrm{po} < 10.01$. \textbf{a} $p$-\ce{Zr3N4}, \textbf{b} $n$-\ce{GaI3Se(CH3)2}, \textbf{c} $n$-\ce{PtC4(NCl3)2}, \textbf{d} $n$-\ce{Zn(CuO2)2}, \textbf{e} $n$-\ce{Mn2CrSbO6}, and \textbf{f} $n$-\ce{Zr3N4}. The shaded regions indicate whether the transport is controlled by the conduction bands ($n$-type) or valence bands ($p$-type).}
    \label{fig:pop-bandstructures}
\end{figure}

In Fig.~\ref{fig:pop-bandstructures} we depict the band structures of the materials with the largest and smallest $T$-dependence of POP limited mobility with $\omega_\mathrm{po} \sim \SI{10}{\tera\hertz}$.
These comprise $p$-\ce{Zr3N4} ($\mu \propto T^{-3.36}$), $n$-\ce{GaI3Se(CH3)2} ($T^{-3.32}$), $n$-\ce{PtC4(NCl3)2} ($T^{-3.13}$), $n$-\ce{Zn(CuO2)2} ($T^{-2.03}$), $n$-\ce{Mn2CrSbO6} ($T^{-1.97}$), and $n$-\ce{Zr3N4} ($T^{-1.95}$).
There are no obvious trends among the band structures.
For example, both large and small $T$ dependencies are seen in band structures with dispersive band edges ($p$- and $n$-type \ce{Zr3N4}, respectively).
Similarly, most of the materials possess degenerate or nearly degenerate band edges with multiple band extrema in the Brillouin zone.

To further investigate the variation in the $T$ dependence of mobility at a single $\omega_\mathrm{po}$, we calculate several band structure features and look for correlations.
These features comprise the conductivity effective mass ($m^{*}_c$), the Seebeck effective mass ($m^{*}_{s}$), the Fermi surface complexity factor\cite{gibbsEffectiveMassFermi2017} ($N_vK^{*} = \left [ m^{*}_{s} / m^{*}_{c} \right ]^{3/2}$), the smallest band width of the bands within \SI{0.3}{\electronvolt} of the band edge ($\Delta E$), and the integrated density of states within \SI{0.3}{\electronvolt} of the band edge ($D_\mathrm{int}$). The conductivity effective mass was calculated according to
\begin{equation}
    m_c^{*}(T,n) = \frac{\sigma(T, n)}{e^2 \tau n},
\end{equation}
where $T$ and $n$ are the temperature and carrier concentration set to \SI{300}{\kelvin} and \SI{e16}{\per\cubic\centi\meter}, respectively, $\sigma$ is the conductivity calculated within the constant relaxation time approximation with a lifetime $\tau = \SI{e-14}{\second}$, and $e$ is the electron charge.
The Seebeck effective mass was calculated using the methodology outlined in Ref.~\citep{gibbsEffectiveMassFermi2017}.

\begin{figure}
    \centering
    \includegraphics[width=\textwidth]{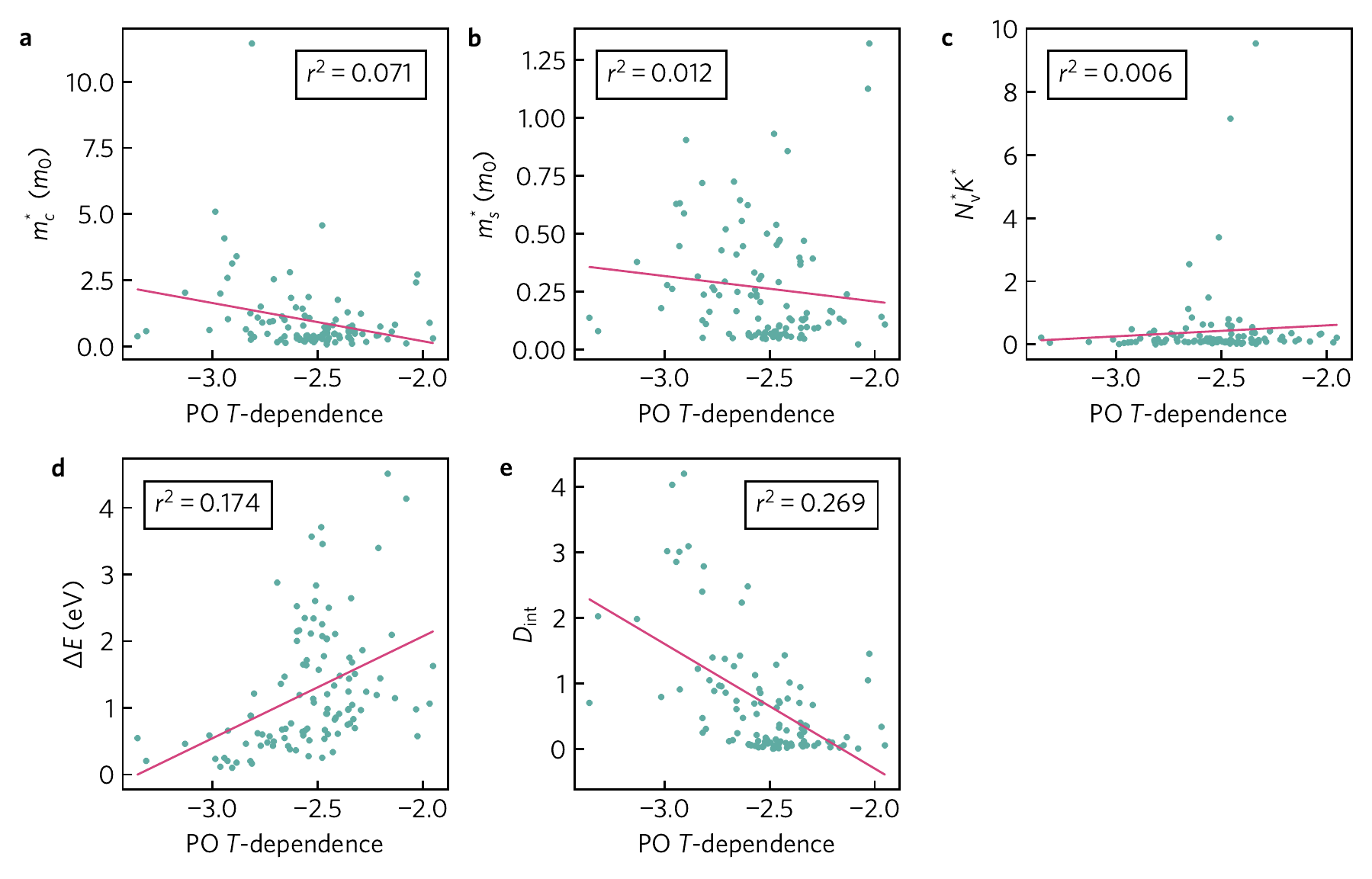}
    \caption{Correlations between band structure features and the temperature of mobility limited by polar optical phonon scattering, for materials where $9.99 < \omega_\mathrm{po} < 10.01$. \textbf{a} conductivity effective mass ($m^{*}_c$), \textbf{b} Seebeck effective mass ($m^{*}_s$), \textbf{c} Fermi surface complexity factor ($N_vK^{*} = \left [ m^{*}_{s} / m^{*}_{c} \right ]^{3/2}$), \textbf{d} the smallest band width of the bands within \SI{0.3}{\electronvolt} of the band edge ($\Delta E$), and \textbf{e} the integrated density of states within \SI{0.3}{\electronvolt} of the band edge ($D_\mathrm{int}$). The pink line is the line of best fit from which the $r^2$ value is calculated.}
    \label{fig:pop-bs-features}
\end{figure}

In Fig.~\ref{fig:pop-bs-features}, we plot the correlation between band structure features and the $T$ dependence of mobility limited by polar optical phonon scattering. 
For each feature, we also calculate the $r^2$ score against the line of best fit as a metric of the correlation.
All features exhibit poor correlation with $T$ dependence.
The integrated density of states (Fig.~\ref{fig:pop-bs-features}e) exhibits the greatest correlation with an $r^2=0.269$ (i.e., the larger number of states is weakly correlated with stronger T-dependence).
The minimum band width within \SI{0.3}{\electronvolt} of the band edge is the second most correlated feature with an $r^2=0.174$ (Fig.~\ref{fig:pop-bs-features}d).
The remaining features ($m^*_c$, $m^*_s$, and Fermi surface complexity factor) all exhibit $r^2$ values below 0.08, indicating essentially no correlation (Fig.~\ref{fig:pop-bs-features}a--c).

\begin{figure}
    \centering
    \includegraphics[width=0.4\textwidth]{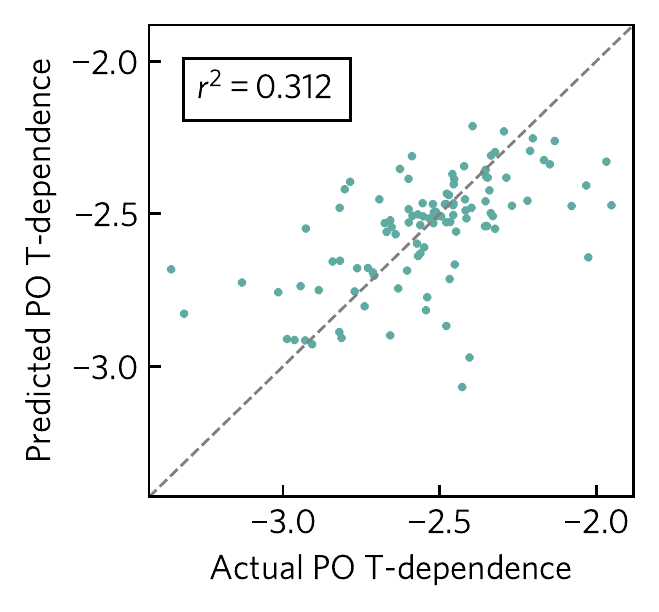}
    \caption{Cross validation predictions of the machine learning model to predict temperature dependence of mobility. Mobility was calculated using only polar optical phonon scattering, for materials where $9.99 < \omega_\mathrm{po} < 10.01$. The dashed line shows the ideal trend.}
     \label{fig:pop-ml}
\end{figure}

Clearly, none of the identified band structure features can account for the variation in $T$ dependence of mobility at a particular $\omega_\mathrm{po}$.
To investigate this further, we trained a machine learning model to assess whether a combination of band structure features can predict the mobility behaviour.
Machine learning was performed using a simple regression model (random forest with 100 estimators).
The dataset comprised the 108 materials where $\SI{9.99}{\tera\hertz} < \omega_\mathrm{po} < \SI{10.01}{\tera\hertz}$ and contained the five features outlined above ($m^*_c$, $m^*_s$, $N_vK^{*}$, $D_\mathrm{int}$, $\Delta E$).
Model performance was evaluated using 5-fold cross validation.
The trained models exhibited essentially no improvement over the single features, with a cross validation $r^{2}$ score of 0.312 (Fig.~\ref{fig:pop-ml}), only marginally larger than for $D_\mathrm{int}$ ($r^2=0.269$).
Accordingly, we posit that the $T$ dependence must be controlled by a complex interplay between band structure effects, wave function overlaps, and the behaviour of the scattering rates across the band structure.

\bibliography{refs}